\documentclass[aps,prd,showkeys,preprint,amsmath,amssymb]{revtex4-1}
\pdfoutput=1

\usepackage{color}
\usepackage{mathrsfs}
\usepackage{graphicx}
\usepackage{array}
\usepackage{latexsym}
\usepackage{enumerate}
\usepackage{slashed}
\usepackage{hyperref}
\usepackage{color}
\def\be{\begin{equation}}
	\def\ee{\end{equation}}

\newcommand{\bear}{\begin{align}}
	\newcommand{\eear}{\end{align}}
\newcommand{\bea}{\begin{align}}
	\newcommand{\eea}{\end{align}}
\newcommand{\nn}{\nonumber}

\def\II{\relax{\rm I\kern-.18em I}}

\def\l{\lambda}
\def\m{\mu}
\def\n{\nu}

\def\sp{\;\;\, ,\;\;\;}

\def\t{\tau}

\def\h{\kappa}
\def\gf{w}
\def\Awf{{A}} %The weight factor in the metric
\begin{document}
\fontsize{12pt}{12pt}\selectfont
{\raggedleft{} CCTP-2016-14, CCQCN-2016-165}

\title{Holographic Photon Production in Heavy Ion Collisions}
%\author{Ioannis Iatrakis \footnote{i.iatrakis@uu.nl}}
%\affiliation{Institute for Theoretical Physics and Center for Extreme Matter and Emergent Phenomena, Utrecht University, Leuvenlaan 4, 3584 CE Utrecht, The Netherlands.}
%\author{Elias Kiritsis \footnote{http://hep.physics.uoc.gr/$\sim$kiritsis/}}
%\affiliation{Crete Center for Theoretical Physics, Institute of Theoretical and Computational Physics, Department of Physics University of Crete, 71003 Heraklion, Greece.}
%\affiliation{Crete Center for Quantum Complexity and Nanotechnology, Department of Physics, University of Crete, 71003 Heraklion, Greece.}
%\affiliation{APC, Univ Paris Diderot, Sorbonne Paris Cit\'e, APC, UMR 7164 CNRS, F-75205 Paris, France.}
%\author{Chun Shen \footnote{chunshen@physics.mcgill.ca}}
%\affiliation{Department of Physics, McGill University, 3600 University Street, Montreal, QC, H3A 2T8, Canada.}
%\author{Di-Lun Yang \footnote{dilunyang@gmail.com}}
%\affiliation{Theoretical Research Division, Nishina Center, RIKEN, Wako, Saitama 351-0198, Japan.}
\author{Ioannis Iatrakis$^{1}$\footnote{i.iatrakis@uu.nl}, Elias Kiritsis$^{2,3,4}$\footnote{http://hep.physics.uoc.gr/$\sim$kiritsis/}, Chun Shen$^5$\footnote{chunshen@physics.mcgill.ca}, Di-Lun Yang$^{6}$\footnote{dilunyang@gmail.com}}

\affiliation{$^1$Institute for Theoretical Physics and Center for Extreme Matter and Emergent Phenomena, Utrecht University, Leuvenlaan 4, 3584 CE Utrecht, The Netherlands.\\
$^2$Crete Center for Theoretical Physics\text{,} Institute of Theoretical and Computational Physics, Department of Physics University of Crete, 71003 Heraklion, Greece.
\\
$^3$Crete Center for Quantum Complexity and Nanotechnology, Department of Physics, University of Crete, 71003 Heraklion, Greece.\\
$^4$APC, Univ Paris Diderot, Sorbonne Paris Cit\'e, APC, UMR 7164 CNRS, F-75205 Paris, France.\\
$^5$Department of Physics, McGill University\text{,} 3600 University Street\text{,} Montreal\text{,} QC\text{,} H3A 2T8\text{,} Canada.\\
$^6$Theoretical Research Division, Nishina Center\text{,} RIKEN, Wako, Saitama 351-0198, Japan.}
%\date{\today}
\begin{abstract}
The thermal-photon emission from strongly coupled gauge theories at finite temperature is calculated using holographic models for QCD in the Veneziano limit (V-QCD).  The emission rates are then embedded in hydrodynamic simulations combined with prompt photons from hard scattering and the thermal photons from hadron gas to analyze the spectra and anisotropic flow of direct photons at RHIC and LHC. The results from different sources responsible for the thermal photons in QGP including the weakly coupled QGP (wQGP) from perturbative calculations, strongly coupled $\mathcal{N}=4$ super Yang-Mills (SYM) plasma (as a benchmark for reference), and Gubser's phenomenological holographic model are then compared. It is found that the direct-photon spectra are enhanced in the strongly coupled scenario compared with the ones in the wQGP, especially at high momenta. Moreover,  both the elliptic flow and triangular flow of direct photons are amplified at high momenta for V-QCD and the SYM plasma. The results are further compared with experimental observations.
\end{abstract}
\keywords{Holography, AdS/QCD, Quark Gluon Plasma, Heavy Ion Collisions}
\maketitle
\tableofcontents
\section{Introduction}

Electromagnetic probes such as photons and dileptons play important roles in both relativistic heavy ion collisions and cosmology \cite{Gale:2009gc,Shen:2016odt}. In heavy ion collisions, considerable amount of direct photons, which do not come from hadron decays, is emitted by the thermal photons emitted from the quark gluon plasma (QGP).

In perturbative quantum chromodynamics (pQCD), the complete leading-order result of photon production in thermal equilibrium incorporating 2 to 2 scattering and collinear emission has been obtained in \cite{Arnold:2001ba,Arnold:2001ms} via the hard-thermal-loop (HTL) resummation, where the perturbative calculation is performed at high temperature with the resummation including dressed propagators and vertices (e.g. the parton self energy computed with soft external momenta as the so called HTL approximation is proportional to $g_s^2T^2$.) in loop diagrams \cite{Bellac:1996}. Recently, the correction from viscous effects has been studied in \cite{Shen:2014nfa}. In addition, the next-to-leading-order calculation was accomplished in \cite{Ghiglieri:2013gia}.

However, due to the strong coupling of the QGP around the deconfinement temperature, perturbative approaches may become invalid for the study of electromagnetic properties of the QGP in Relativistic Heavy Ion Collider (RHIC) and Large Hadron Collider (LHC). Although the electric conductivity of the QGP has been recently obtained from lattice simulations \cite{Amato:2013naa,AartsAlltonAmatoEtAl2015}, the non-perturbative computation of full spectra of photon or dilepton production in sQGP is still formidable.

The gauge/gravity duality, which connects the strongly coupled gauge theories and classical supergravity, provides an alternative way to tackle non-perturbative problems \cite{Maldacena:1997re,Gubser:1998bc,Witten:1998qj}. It has been widely applied to investigate the heavy ion collisions (see \cite{CasalderreySolana:2011us} and the references therein). In holography, the study of thermal-photon production and dilepton production from a strongly coupled $\mathcal{N}=4$ SYM plasma was initiated by \cite{CaronHuot:2006te}.

Similar calculations were performed in top-down models that include quarks in the fundamental representation by adding flavor branes to the SYM theory, \cite{Mateos:2007yp,Parnachev:2006ev}. In addition, the photoemission at intermediate t'Hooft coupling was analyzed in \cite{Hassanain:2011ce,Hassanain:2012uj}. Further studies incorporated strong magnetic fields \cite{Mamo:2013efa,Yee:2013qma,Muller:2013ila} and pressure anisotropy \cite{Patino:2012py,Wu:2013qja,Jahnke:2013rca}, which are relevant to electromagnetic probes generated at the early stage of heavy ion collisions. Moreover, the photon production in the non-equilibrium state in holography were presented in \cite{Baier:2012ax,Baier:2012tc,Steineder:2012si,Steineder:2013ana,Mamo:2014ema}.

On the other hand, recent measurements of large anisotropic flow of direct photons comparable to the hadron flow in RHIC \cite{Adare:2011zr} and LHC \cite{Lohner:2012ct} lead to a new puzzle. It is generally believed that electromagnetic probes only record the local information in heavy ion collisions. Due to the weak electromagnetic coupling, they barely interact with the QGP after they are produced. Some of the recent studies indicate that the interaction between electromagnetic probes and the QGP could be enhanced compared with the case in the quantum electrodynamics (QED) plasma \cite{Muller:2015maa,Yang:2015bva}, whereas such an enhanced interaction should be negligible due to short lifetime of the QGP.

 In general, since thermal photons inherit the flow from their sources, it is expected that the anisotropic flow of direct photons at intermediate energy around, $1-3$ GeV, should be smaller than the flow of hadrons given that a large portion of direct photons are generated in early times with small momentum anisotropy. Theoretically, the state-of-the-art simulation incorporating the emission of prompt photons from hard interactions in early times before thermalization and of thermal photons produced
in the weakly coupled QGP with HTL resummation and the rate from hadron gas along with the plasma expansion dictated by viscous
hydrodynamics, under-predicts the elliptic flow compared with the experimental measurements \cite{Dion:2011pp,Shen:2013cca}.
 
On the other hand, a variety of non-conventional mechanisms have been proposed to address the puzzle \cite{Bzdak:2012fr,Fukushima:2012fg,Muller:2013ila,Linnyk:2013wma,McLerran:2014hza,Gale:2014dfa,Monnai:2014kqa,Linnyk:2015tha,McLerran:2015mda}. A more recent study, \cite{Shen:2013cca}, which incorporates the IP-Glasma initial states and the influence from both shear and bulk viscosities with updated photon emission rates, improves the agreement with experimental observations \cite{PaquetShenDenicolEtAl2016}. In addition, the effects from late equilibration of (anti-)quarks in the QGP evolution was investigated in \cite{Vovchenko:2016ijt}, which yield mild suppression of the spectrum but an enhancement of flow for direct photons.

Despite the validity at the RHIC and LHC temperature, the leading-order emission rate of thermal photons from the weakly coupled QGP is applied \cite{Arnold:2001ba,Arnold:2001ms} to describe the thermal radiation from the deconfined phase in the simulations due to the difficulty in tackling photon emission from the sQGP.

In this paper we employ two bottom-up holographic models to model the sQGP, both of which break conformal invariance and match several  properties of QCD at finite temperature as calculated from lattice simulations. In holography, the running of the coupling can be characterized by a bulk scalar field (dual to the operator driving the RG flow) with a corresponding potential in the
gravity dual. This type of models may be regarded as effective theories of QCD in
the infrared (IR) regime. Although the validity  of such models may not be rigorously justified, they quantitatively agree with many properties of QCD and  provide an alternative route to probe the direct-photon problem and improve our understandings for the present tension between theories and experiments.

One of the models we employ was introduced by Gubser and Nellore \cite{Gubser:2008ny,Gubser:2008yx}.
In \cite{Finazzo:2013efa}, the model was  utilized to fit the lattice data for the electric charge susceptibility and qualitatively describes the electric conductivity from lattice data near the deconfinement phase transition \cite{Greif:2014oia}. Furthermore, the thermal-photon production in such a model was analyzed and was briefly reported in \cite{Yang:2015bva}. Hereafter we will call this model of GN model.

The other model we will use is  V-QCD. This is a more sophisticated holographic model and it is based on the improved holographic QCD formalism \cite{Gursoy:2007cb,Gursoy:2007er,Gursoy:2008bu,Gursoy:2008za}.
The flavor degrees of freedom are included by adding $N_f$ brane-antibrane pairs with a bulk tachyon field that describes the spontaneous breaking of chiral symmetry
\cite{Bigazzi:2005md,Casero:2007ae,Iatrakis:2010jb,Iatrakis:2010zf}.

The framework of V-QCD works  in the Veneziano limit
\be
N_{c,f}\rightarrow \infty\;\;\;,\;\;\; x=N_f/N_c=\text{fixed}
\ee
and includes the backreaction of flavor on color \cite{Jarvinen:2011qe}.
Its phase structure at zero temperature, finite temperature and finite temperature and density was analyzed in a series of works \cite{Alho:2012mh, Arean:2012mq, Arean:2013tja,  Arean:2013lta, Alho:2013hsa, Iatrakis:2014txa,  Alho:2015zua, Jarvinen:2015qaa, Iatrakis:2015rga}. In particular, several transport coefficients and flavor current two point functions were analysed at finite temperature and chemical potential in \cite{Iatrakis:2014txa}.
V-QCD contains in particular the dynamics of chiral symmetry breaking that is absent from the GN model.   

The photon-emission rates from these holographic models are then convoluted with the medium evolution. Furthermore, the contributions from prompt photons and thermal photons from hadron gas are incorporated to compute both the spectra and flow of direct photons in both RHIC and LHC energies.
Here, we summarize our findings in short:

\begin{enumerate}
	\item
	Holographic photon production results in enhanced spectra of direct-photons especially for direct photons with high transverse (perpendicular to the beam direction in experiments) momenta $p_T$.
	\item
	The anisotropic flow of direct photons is suppressed in both holographic models compared to pQCD with HTL resummation at low $p_T$, while V-QCD enhances their $v_n$ at high $p_T$.
	\item
	Holographic models improve the agreement with experimental observations in spectra, but still underestimate the flow.
	\item
	In small collision systems, holographic models give rise to enhancements in both spectra and flow.
\end{enumerate}

Based on the findings, we also present  our conclusions below.

\begin{enumerate}
	\item
	The hadronic contribution for thermal photons plays a central role for direct-photon flow in low $p_T$ as concluded in previous studies.
	\item
	For $p_T \in [2, 4]$ GeV, thermal photons from the QGP phase are responsible for the direct photon anisotropic flow. In this case, the direct-photon flow can be further increased when the emission rate is more prominent at low temperature and the yield is more dominant than prompt-photon contribution for the corresponding model.
	\item
	In small collision systems, larger spectra of thermal photons from the QGP phase simultaneously yield larger flow for direct photons.  	
\end{enumerate}

The paper is organized as following. In section \ref{sec_vqcdgen}, we introduce basic ideas about the constructions of holographic QCD from bottom-up approaches. In section \ref{sec_holography_gen}, we elaborate the holographic models we will apply for photon production. We then show the technical details for the computations of photon emission rates and compare the results from distinct models in section \ref{sec_photon_emission}. In section \ref{sec_spec_and_flow}, we first expatiate the setup of the full simulation for the direct-photon production in heavy ion collisions. Then we present the spectra and elliptic flow in RHIC and LHC, in which we further make discussions and comparisons between the results and experimental observations. Finally, we make concluding remarks in section \ref{sec_con}.

\section{Holographic QCD}
\label{sec_vqcdgen}
%%%%%%%%%%%%%%%%%%%%%%%%%%%%%%%%%%%%%%%%%%%%%%%%%%%%%%%%%%%%%%%%%%

There are several approaches to the holographic study of strongly coupled QCD. The deconfined phase of the theory at intermediate temperatures already bears some similarities with the ${\mathcal N}=4$ SYM theory. An effort to describe holographically QCD features in more detail, led to the development of phenomenological bottom-up holographic models, \cite{Gursoy:2007cb,Gursoy:2007er,Kiritsis2009,GursoyKiritsisMazzantiEtAl2011} and \cite{Gubser:2008ny}.

Current holographic approaches to the QCD problem with light quarks are carried in the context of the probe approximation
whereby the light quarks are inserted as spectator D-branes in the limit where $N_f\ll N_c$. The Veneziano limit consists of
addressing the double limit $N_{f,c}\rightarrow \infty$ but fixed $\lambda=g^2N_c$ and $x=N_f/N_c$ in the context of  holographic models or V-QCD.  In the vacuum V-QCD exhibits chiral symmetry breaking for $x\leq 4$, restores chiral symmetry for  $4\leq x\leq 5.5$, \cite{Jarvinen:2011qe}. Beyond the Banks-Zaks point or $x>5.5$, \cite{Banks:1981nn}, the theory becomes QED-like.

The bulk action describing the system
\be
S = S_g + S_f + S_a \,.
\ee
$S_g$ is the gluon action which is the same as the improved-holographic QCD (ihQCD) action, \cite{GursoyKiritsisMazzantiEtAl2011}. $S_f$ is the brane-antibrane action which was introduced in \cite{Sen2005}, and was proposed as a low energy effective action for the holographic meson sector in \cite{Bigazzi:2005md, Casero:2007ae, Iatrakis:2010zf, Iatrakis:2010jb}. The last part, $S_a$, is the action of the CP-odd sector which contains the coupling of the flavor-singlet mesons to the axion, which comes from the closed string sector, \cite{Bigazzi:2005md, Casero:2007ae, Arean:2013tja}. 

The full form of the first two parts of the action will be presented below. Since, we currently do not study the flavor-singlet CP-odd excitations and the vacuum state is CP-even the last term is not discussed further. To study the excitation spectrum we expand the above equations to quadratic order around the vacuum solution.

%%%%%%%%%%%%%%%%%%%%%%%%%%%%%%%%%%%%%%%%%%%%%%%%%%%%%%%%%%
\subsection{The glue sector}
\label{sec:VQCDglue}

The glue part of QCD at large $N_c$ and strong t' Hooft coupling is described holographically by a two-derivative effective action with bulk fields that correspond to the lowest dimension operators of the theory. Those include the metric and the dilaton. According to gauge/gravity duality, the metric and the dilaton are dual to the energy-momentum and the ${\mathbb Tr} F^2$ operator, respectively.  The following gluonic action is used to describe the dynamics

\be
S_g= M^3 N_c^2 \int d^5x \ \sqrt{-g}\left( R- {4 \over 3}{
	(\partial \lambda)^2 \over \lambda^2} + V_g(\lambda) \right) \, .
\label{glueact}
\ee
$M^3=1/(16\pi G_5N_c^2)$ is the 5-dimensional Plank mass and $\l=e^\phi$ is the exponential of the bulk dilaton with the corresponding potential $V_g$. $\lambda$  is interpreted as the holographic t' Hooft coupling. The Ansatz for the finite temperature vacuum solution is

\be
ds^2=e^{2 \Awf(r)} \left( -f(r)\,dt^2+dx_{3}^2+{dr^2 \over f(r)} \right) \, \, , \, \,\, \lambda=\lambda(r) ,
\label{bame}
\ee
where the conformal factor $e^{2 A}$ is identified as the energy scale in field theory and $f(r)$ is the black hole factor. The choice of the potential $V_g(\phi)$ depends on the details of each holographic model. In this work, we will consider three different models in order to compare their results for the photon production in Heavy-Ion collisions. Our models include the prototype AdS-Schwarzschild background, where $V_g(\lambda)$ is constant, the Improved Holographic QCD model, \cite{GursoyKiritsisMazzantiEtAl2011,Gursoy:2008bu,Gursoy:2008za,Gursoy:2009jd}, and the GN model, \cite{Gubser:2008ny}.

%%%%%%%%%%%%%%%%%%%%%%%%%%%%%%%%%%%%%%%%%%%%%%%%%%%%%%%%%%
\subsection{The flavor sector} \label{sec:VQCDflavor}

In the current work, we are interested in describing certain electromagnetic properties of the Quark Gluon Plasma, hence we have to include the dynamics of the electric current of the plasma. The electric current is an operator in the flavor sector of the field theory, hence one has to include flavors in the holographic picture. In case of ${\mathcal N}=4$ SYM, the $U(1)$ electric current is taken to be a conserved current corresponding to the $U(1)$ subgroup of the $SU(4)$ R-symmetry of ${\mathcal N}=4$ SYM.

In the GN model, the $U(1)$ flavor current is described phenomenologically by including a bulk $U(1)$ gauge field, with the standard kinetic term multiplied by a dilaton dependent potential.

In V-QCD, a more thorough approach is followed since the dynamics of the flavor sector (both in the confined and the deconfined phase) is described by introducing appropriate bulk fields for all the low dimension operators of the theory. Those include the quark condensate 
$\bar\psi_R\, \psi_L$,  as well as the left and right flavor currents $\bar\psi_L\,\gamma_{\mu}\,\psi_L$ and $\bar\psi_R\,\gamma_{\mu}\,\psi_R$.

To describe the dynamics of the above operators and couple them to the glue part of the theory, we effectively consider $N_f$ pairs of $\bar D_4$ and $D_4$ coincident branes in the 5 dimensional bulk spacetime. The dynamics of such a a brane system in flat space-time was described in \cite{Sen2005}. The lowest lying fields of the system include a complex scalar field $T_{I\bar J}$ (dual to the quark condensate ) that transforms in the $(N_f,\bar N_f)$ of the flavor group $SU(N_f)_L\times SU(N_f)_R$. It is the famous tachyon of brane-antibrane systems in flat space and it also a tachyon in the asymptotically AdS space relevant for holography. The tachyon instability translates into the fact that the quark mass operator is a relevant operator in QFT.
 
There are also two $U(N_f)$ gauge fields, $A^{L}_{\m}$ and $A^{R}_{\m}$, dual to the left and right flavor currents.
The dynamics of both  the tachyon and the gauge fields is determined  by the action, which was proposed by Sen.
The condensation of the tachyon   describes the chiral symmetry breaking in  the vacuum of the theory.
In the current article, we are interested in the finite temperature, deconfined state of the model in which chiral symmetry is restored.
This is described by a black hole that has dilaton hair but no tachyon hair, \cite{Alho:2012mh}. %\EK{Cite the paper on finite temperature and V-QCD here}

\section{The different holographic  models}\label{sec_holography_gen}

\subsection{${\mathcal N}=4$ Super-Yang-Mills}

The glue part of the action is the same as (\ref{glueact}) with the dilaton set to zero and $V_g={12\over \ell^2}$, where $\ell$ is the AdS radius. The background solution at finite temperature is given by (\ref{bame}), with

\be
\Awf(r)=\log\left( {\ell \over r}\right) \,\, ,\,\,\, f(r)=1-{r^4 \over r_h^4} \,.
\ee
In our discussion, we will omit the 5-dimensional sphere which is a part of the background metric because it will not play any role.

 The flavor part of the action describing the $U(1)$ bulk gauge field dual to the photon current is simply 
\be
S=-{1\over 4 g_e^2} \int d^5x \sqrt{-g} F_{MN} F^{MN}\sp g_e^2=16 \pi^2 \ell/N_c^2\;.
\ee

\subsection{The Gubser and Nellore model}

In the case of the GN model, the dilaton potential is chosen in order for the model to reproduce the the equation of state of the Quark Gluon Plasma near $T_c$\footnote{Note the this model is only capable of describing the deconfined phase. At zero temperature it is different from QCD as it is not confining and does not have a mass gap.}, \cite{Finazzo:2013efa} and \cite{Yang:2015bva}.  $V_g(\lambda)$ reads

\be
V_g(\lambda)=6 (\lambda^{\gamma} + \lambda^{-\gamma})-b_2 \log^2 \lambda-b_4 \log^4 \lambda-b_6 \log^6 \lambda \, ,
\ee
where $\gamma=0.606$, $b_2=0.703$, $b_4=-0.12$ and $b_6=0.0044$ and $\ell=1$. The background metric and dilaton are found numerically by integrating Einstein's equations.

The flavor sector here contains also a $U(1)$ gauge field dual to the photon which is coupled to gravity by the following action

\be
S_f=-M^3 N_c^2 \int d^5x \sqrt{-g}{V_e(\lambda) \over 4} F^{MN}F_{MN} \, ,
\ee
where the potential of the coupling of the dilaton to the gauge field is taken to be, \cite{Yang:2015bva},

\be
V_e(\lambda)=2 g_e^2 {1 \over \lambda^{a_1}+\lambda^{-a_1}} \, ,
\label{gnmodel}
\ee
where $g_e^2$, is a dimensionless constant which is matched to the overall amplitude of the relevant observables, and  $a_1=0.4$. This is a phenomenological choice of the potential which matches the lattice results for the electric susceptibility of the QGP for $T<1.5 T_c$. In \cite{Yang:2015bva}, $g_e$ is chosen to saturate the electric conductivity of the SYM plasma at $T=3.5 T_c$. In the subsequent study \cite{Finazzo:2015xwa} with the inclusion of chemical potentials, a different choice of $g_e$ is considered, which results in smaller electric conductivity and photon-emission rates in the zero chemical potential.

\subsection{V-QCD}

The bulk fields corresponding to the lowest dimension meson operators  include two vector gauge fields and a complex scalar, the tachyon. Their effective holographic action was proposed in \cite{CaseroKiritsisParedes2007,Iatrakis:2010zf,Iatrakis:2010jb},
%\textcolor{blue}{(To avoid confusion with temperature, I hereafter write tachyon T as $\bar{T}$)}
\be
%\textcolor{red}{
S_f= - \frac{1}{2} M^3 N_c  {\mathbb Tr} \int d^4x\, dr\,
\left(V_f(\l,\bar{T}^\dagger \bar{T})\sqrt{-\det {\bf A}_L}+V_f(\l, \bar{T}^\dagger \bar{T})\sqrt{-\det {\bf A}_R}\right)\,,
%}
\label{generalact}
\ee
where the definitions of ${\bf A}_{L/R}$ are
\begin{align}
	{\bf A}_{L\,MN} &=g_{MN} + \gf(\l,\bar{T}) F^{(L)}_{MN}
	+ {\h(\l,\bar{T}) \over 2 } \left[(D_M \bar{T})^\dagger (D_N \bar{T})+
	(D_N \bar{T})^\dagger (D_M \bar{T})\right] \,,\nonumber\\
	{\bf A}_{R\,MN} &=g_{MN} + \gf(\l,\bar{T}) F^{(R)}_{MN}
	+ {\h(\l,\bar{T}) \over 2 } \left[(D_M \bar{T}) (D_N \bar{T})^\dagger+
	(D_N \bar{T}) (D_M \bar{T})^\dagger\right] \, , \nn \\
	D_M \bar{T} &= \partial_M \bar{T} + i  \bar{T} A_M^L- i A_M^R \bar{T}\,,
	\label{Senaction}
\end{align}
where $g_{MN}$ is the induced metric on the brane-antibrane pair. The fields  $A_{L}$, $A_{R}$ as well as the tachyon field $\bar{T}$ are $N_f \times N_f$ matrices in the flavor space. ${\bf A}_{L/R}$. The Plank mass, which appears as an overall factor in front of both $S_g$ and $S_f$,  is fixed by requiring the pressure of the system  to approach the large temperature limit of free non-interacting fermions and bosons. This fixes $(M \ell)^3 = (1+7 x/4)/ 45 \pi^2$ \cite{Alho:2012mh}, and the AdS radius in the presence of backreacting flavors is
$\ell^3 = \ell_0^3 \left( 1+{7 x \over 4} \right) $.  The quarks are taken to have the same mass, so the tachyon field is proportional to the unit matrix in flavor space, $\bar{T} =\tau(r) \mathbb{I}_{N_f}$. The field $\tau(r)$ is real.

The near-boundary expansion of the tachyon field matches the UV running of t' Hooft coupling and the anomalous dimension of the quark mass operator \cite{Jarvinen:2011qe}. The tachyon close to the boundary $r\to 0$ is

\be
\tau(r)= m_q r(-\log \Lambda r)^{- \gamma}+\langle {\bar q} q \rangle r^3 (-\log \Lambda r)^{\gamma}+\cdots \, ,
\label{tauuv}
\ee
where the power  $\gamma$ is matched to the coefficients of the anomalous dimension of ${\bar q} q$.
Here $m_q$ and $\langle {\bar q} q \rangle$ denote the quark mass and quark condensate, respectively. In the IR, the tachyon field diverges and the tachyon potential vanishes as it is argued in \cite{Sen2005}. As it is shown in \cite{CaseroKiritsisParedes2007,Iatrakis:2010zf,Iatrakis:2010jb} brane - antibrane condensation in confining backgrounds leads to chiral symmetry breaking. The form of the tachyon potential that we use is
\be
V_f(\l,\bar T \bar T^\dagger)=V_{f0}(\l) e^{- a(\l) \bar T \bar T^\dagger} \, .
\label{tachpot}
\ee
The rest of the potentials $\kappa(\lambda)$ and $w(\lambda)$ are taken to be independent of $T$ and  have an analytic expansion close to the boundary in terms of $\lambda$. Chiral symmetry breaking, thermodynamics and the meson spectra constrain their IR asymptotics, \cite{Arean:2013tja} potentials $V_{f0}(\lambda)$,  $\kappa (\lambda)$  and $a(\l)$ are given by

\begin{eqnarray}
	\label{Vf0SB}
	%\hspace{-1cm}
	V_{f0}& =& {12\over \ell^2}\biggl[{\ell^2\over\ell_0^2}
	-1+{8\over27}\biggl(11{\ell^2\over\ell_0^2}-11+2x_f\biggr)\lambda
	\nn \\ && % \hspace{-1cm}
	+{1\over729}\biggl(4619{\ell^2\over \ell_0^2}-4619+1714x - 92x^2\biggr)\lambda^2\biggr] \, , \nn \\
	%\equiv W_0+W_1\l+W_2\l^2.
	\kappa(\l) &=& {[1+\ln(1+\l)]^{1/2}\over[1+\frac{3}{4}(\frac{115-16x }{27}+{1\over 2})\l]^{4/3}} \quad a(\l)=\frac{3}{2 \, \ell^2} \, .
\end{eqnarray}

It should be noted that the potentials in IR are remarkably close to the non-critical string theory values on flat space-time, except form the logarithmic corrections, see \cite{Arean:2012mq,Arean:2013lta,Arean:2013tja}.
In our present work, we choose $w(\l)$, in such a way that the holographic electric conductivity of the QGP in the deconfined phase.

The boundary of the bulk space-time is taken to be at $r=0$ and the field-asymptotics are modified from the usual AdS asymptotics such that they reproduce the perturbative running of t' Hooft coupling.  Hence in the UV,
the scale factor of the metric $A$ and the 't Hooft coupling $\l$ are 
\be
A \sim \ln \left( { \ell \over r} \right)+{4 \over 9 \log (\Lambda r)}+\ldots \,\, ,\,\,\,  \lambda \sim  { 1 \over  \log (\Lambda r)}+\ldots  \,, \,\,  r \to 0 \,,
\label{uvasal}
\ee
where  $\ell$ is the AdS radius and $\Lambda$ is the UV scale of the theory.
The dilaton potential is chosen such that it reproduces the perturbative running of t' Hooft coupling in the UV. Even if QCD in the UV is not expected to have a dual description in terms of a gravity theory, since it is weakly coupled, the UV asymptotics of our model provide the correct UV-boundary conditions for the IR description of the system. The requirement of confinement, gapped and discrete glueball spectrum as well as linear Regge trajectories fix the IR behavior of the potential to $V(\lambda) \sim \lambda^{4/3} \sqrt{\ln \lambda}$ as $\l\to\infty$.  Considering a simple interpolation of the asymptotic forms of the potential in the two regions we obtain
\be
V_g(\lambda)={12\over \ell_0^2}\biggl[1+{88\lambda\over27}+{4619\lambda^2
	\over 729}{\sqrt{1+\ln(1+\lambda)}\over(1+\lambda)^{2/3}}\biggr]\, ,
\label{VfSB}
\ee
where $\ell_0$ is the AdS radius in case of no backreacting flavors.

\subsection{The vector field action}

To calculate correlation functions of the electric current in QGP, we focus on the vector gauge field in the bulk, which is defined as $V_M = \frac{A_M^L + A_M^R}{2}\, $, in terms of the left and right gauge fields, \cite{Iatrakis:2014txa}.  We expand the action, (\ref{generalact}), to quadratic order in $V_M$

\be
\begin{split}
	S_V & =- M^3 \, N_c \,  {\mathbb Tr} \, \int d^4 x \, dr  V_f(\l,\t) \sqrt{-\mathrm{det} \, g} \sqrt{ \mathrm{det} (\delta^M_N +w(\lambda) g^{MR}V_{RN} )} \\
	& =-M^3 \, N_c \, {\mathbb Tr} \, \int d^4 x \, dr  \sqrt{-\mathrm{det} \, g } \left(V_f(\l,\t) + {V_{f}(\l,\t)  w(\l)^2 \over 4} F^{MN} \, F_{MN} + \ldots \right)
\end{split}
\ee

In the chirally symmetric phase of the model the tachyon is zero, $\bar{T}=0$, hence the above potentials depend only on the dilaton. Hence, in this case the coupling function of the dilaton to the gauge field reads
\be
V_{e}(\lambda)=V_{f0}(\l)  w(\l)^2 \, .
\label{vqcdve}
\ee

In the current paper, we make two different choices of the potential $w(\l)$ and explore its consequences to the final photon spectrum and it's flow.

\begin{enumerate}
	\item
	$w_1(\lambda)={1 -{3 \over 5} \tanh (\lambda-\lambda_c) \over (1+ \lambda)^{4/3} },$
	\item
	$w_2(\lambda)={1 -{3 \over 5} \tanh (\lambda-\lambda_c) \over (1+ \lambda)^{4/3} }+{1\over 30 \lambda^3}.$
\end{enumerate}

The two choices are motivated by the exploration of the holographic possibilities. The reason is as follows: Lattice calculations provide the electrical conductivity of QCD in a finite energy window, namely $T_c<T<2T_c$.
This constraints directly $w(\l)$ in a finite range of couplings $\l_c<\lambda<\lambda_{cc}$. $w(\l)$ for $\lambda>\l_c$ is not constrained from current lattice results.

The two distinct parametrizations above, although both in agreement with lattice results for $\lambda<\l_{cc}$   are different for $\l>\l_{cc}$.
In the second case, $w_2(\l)$ increases faster than $w_1(\lambda)$, for low values of $\lambda$ or equivalently high temperature\footnote{$\lambda_{cc}$  is the horizon value of the dilaton at $T=2Tc$ while $\l_c$ is the horizon value of the dilaton at the deconfinement transition $(T=T_c)$. This value is determined by calculating the phase diagram of the model, and for the above choice of potentials it reads $\lambda_c=1.691$.} This choice was made in order to explore the consequences of a high UV-contribution in the final photon spectrum. This will be explained in more detail in the following.

%%%%%%%%%%%%%%%%%%%%%%%%%%%%%%%%%%%%%%%%%%%%%%%%%%%%%%%%%%
\subsection{The vector fluctuations}
\label{app:NSvmesons}

%\EK{Here you do the calculation only for GN model. You should also add the calculation for V-QCD} \\
%\textcolor{green}{Yiannis: The following  calculation is the same for both models since we study the quadratic action of the vector fluctuations. The only difference between the two models is the background and the function $V_e(\l)$. For GN $V_e$ is given by  \ref{gnmodel} and for V-QCD  by \ref{vqcdve}. The same holds for the next section.}

We calculate the action and the equations of motion for the fluctuation of the vector gauge field in the deconfined background, where chiral symmetry is restored ($\tau(r)=0$),  \cite{Iatrakis:2014txa}. The calculation is common for the different models which we consider, since they only differ in the background solutions and the choice of the potential $V_e(\l)$. The quadratic action for the fluctuation of the vector field in the $V_r=0$ gauge is

\be
\begin{split}
	S_V &= -  {1\over2}\, M^3 N_c\,  {\mathbb Tr} \, \int d^4x\, dr
	V_e(\l) \,e^{A(r)} \\
	& \left[ \frac12\, V_{ij}V^{ij} -\frac12 \, {1\over f(r)^2}\, V_{i0}V^{i0}
	+f(r)  \partial_r V_i \partial_r V^{i} -\partial_r V_0 \partial_r V^{0}
	\right]\, ,
\end{split}
\label{vectoracti1m}
\ee
where   $V_{ij}=\partial_i V_j-\partial_j V_i$, and the trace is over the flavor indices. We define the Fourier transform of the vector field as

\be
V_\mu (t,{\bf x},r) =\int {d^4 k \over (2 \pi)^4} e^{-i \omega t+i {\bf k x}}  V_\mu (r,\omega,k)\,.
\label{fd}
\ee
Without  loss of generality, we may take $k_3=k\neq 0$ and $k_0=-\omega$. Then, the two decoupled equations for the transverse $V_i^{\bot}$ and the longitudinal gauge invariant field ${E_L}= k V_0 +\omega V_3$ are

\begin{align}
	& -\partial_r \biggl(  e^{A(r)} \, f(r) \,   V_e(\l)  \, \partial_r \, V^\bot_i(r)  \biggr) % \, \nonumber \\
	%&
	+ e^{A(r)} \,  V_e(\l) \, \biggl( k^2 -  { \omega^2 \over f(r) } \biggr)   V^\bot_i(r) \,=0 \, ,
	\label{tranveceq1}
\end{align}

\begin{align}
	&E_L(r)'' - \left( \partial_r \log  (V_e(\l) e^{A(r)} )   +{\omega^2 \over \omega^2 - k^2 f(r)} \partial_r  \log f(r)   \right) E_L(r)' \nn \\
	&-{1 \over f(r)^2} \left( \omega^2 - k^2 f(r) \right) E_L(r)=0 \, .
	\label{longveceq1}
\end{align}

For ${\bf k}=0$, the equations of $V^\bot$ and $E_L$ reduce to the same equation

\be
\frac{1}{ V_e(\l) \, e^{\Awf}\, f^{-1}}
\partial_r \left( V_e(\l) \,e^{\Awf}\, f \, \partial_r \psi_V \right)
+\, \omega^2  \,\psi_V = 0 \, ,
\label{vectoreom1m}
\ee
the vector field in the bulk is written as $V_i^{\bot}=V_3=\psi_V$, where $i=1,2$ denotes the transverse coordinates.

%%%%%%%%%%%%%%%%%%%%%%%%%%%%%%%%%%%%%%%%%%%%%%%%%%%%%%%%%%%%%%%%%%%%%%%%%%%%%%%%%%

\section{Thermal-Photon Production in Holography}\label{sec_photon_emission}
\subsection{The Spectral function}
\label{specfun}

The retarded correlator in momentum space is defined as

\be
G_{\mu\nu}^{ab \,R}(k)=\int d^4(x-y)e^{ik\cdot(x-y)} \theta(x^0-y^0) \langle [ J^a_{\mu}(x), J^b_{\nu}(y) ] \rangle
\ee
where $a,b$ are the $SU(N_f)$ indices. The correlator of the vector and transverse axial-vector current are proportional to $P_{\mu\nu}= \eta_{\mu\nu}-{k_\mu k_\nu \over k^2}$, $G_{\mu\nu}^{ab \, R}=P_{\mu\nu} \Pi^{ab}(k^2)$. In the thermal states that we consider Lorentz symmetry is broken and only rotational symmetry is left. Then, the projector is split in transverse and longitudinal parts with respect to the spatial momentum ${\bf k}$, $P_{\mu\nu} = P_{\mu\nu}^T + P_{\mu\nu}^L$, where the transverse part is defined as

\begin{align}
P^{T}_{00}=0\, , \,\, P^{T}_{0i}=0 \, ,\,\, P^T_{ij}=\delta_{ij}-{k_i k_j \over k^2} \,, \,\,
\end{align}
and $P^L_{\m\n}=P_{\m\n}-P^T_{\m\n}$. The Green's function reads
%\textcolor{blue}{(better to add the definitions of $P_{\mu\nu}^T$ and $P_{\mu\nu}^L$.)}

\be
G_{\mu\nu}^{ab \, R}= P_{\mu\nu}^T \Pi_T^{ab}(k^2) +P_{\mu\nu}^L \Pi_L^{ab}(k^2) \,.
\ee

%\EK{Also below you seem to calculate in the GN Model only. Please include the equivalent expressions for VQCD}

In thermal equilibrium, the photon production is given by the light-like correlator (see e.g.\cite{Bellac:1996}),
\begin{eqnarray}
d\Gamma=-\frac{d^3k}{(2\pi)^3}\frac{e^2n_b(|{\bf k}|)}{|{\bf k}|}\text{Im}\left[\text{tr}\left(\eta^{\mu\nu}G_{\mu\nu}^{ab \, R}\right)\right]_{k^0=|{\bf k}|},
\end{eqnarray}
where $\Gamma$ denotes the number of photons emitted per unit time per unit volume and $n_b(|{\bf k}|)$ denotes the thermal distribution function for Bosons. Here the trace is taken with respect to flavors.
We are thus interested in the transverse part of the correlator for vector fluctuations.
We Fourier decompose the vector field, Eq.(\ref{fd}), choose $k_\m=(-\omega,0,0,k_3)$ with $|k_3|=k$ and use the variable $E_L=\omega V_3 + k_3 V_0$. Using the equations of motion (\ref{tranveceq1}) and (\ref{longveceq1}), we find from (\ref{vectoracti1m}) the on-shell action

\begin{eqnarray}
	S_V =   { M^3 N_c N_f \over 2} {\delta^{ab} \over 2 N_f} \int &&  {d^4k \over (2 \pi)^4} \Big[
	V_e(\l) \, f\,e^{\Awf}   \nn \\
	&& \Big(   {{\cal E}^a(-k) {\cal E}^b(k) \psi_E(r) \partial_r \psi_E(r) \over \omega^2 -f\,k^2}   + {\cal V}^a_i(-k) {\cal V}^b_i(k)  \psi_V(r) \partial_r \psi_V(r) \Big) \Big]_{r=\epsilon}\, .
	\label{vectoractionsh}
\end{eqnarray}
where 
\be
V_i (k, r) =  \psi_V(r) \, {\cal V}^a_i (k) t^a \sp E_L (k, r) =  \psi_E(r) \, {\cal E}^a (k) t^a
 \ee
with $t^a$ being SU($N_f$) flavor matrices. Taking the second derivative with respect to the sources ${\cal V}_i$ we find the transverse part of the retarded two-point function

\be
G_{ij}^{ab \, R}(\omega,k)= {\delta^{ab} \over 2 N_f} \left(\delta_{ij} - {k_i k_j \over {\bf k}^2 }\right) \Pi_T(\omega,k) \, .
\ee
$\Pi_T$ is calculated from the bulk on-shell action and is normalized in order to be the same as as in the flavor singlet case. It is found directly from (\ref{vectoractionsh})

\be
\Pi_T= -M^{3} N_c N_f   \left( V_e(\l) \, f\,e^{\Awf} \, \mathrm{Im} \left( \psi_V \partial_r \psi_V \right) \right|_{r=\epsilon} \, ,
\label{spectdensfin}
\ee
where $\psi_V$ is the solution of \eqref{vectoreom1m} with infalling boundary condition on the horizon and $\psi_{V}(\epsilon)=1$ at the boundary. The spectral density is defined as

\be
\rho=- 2\, \mathrm{Im} G_{ii}^{R}= -6\, \mathrm{Im} \Pi_T\, ,
\ee
with the spatial indices being contracted. The spectral density can be calculated using the membrane paradigm by redefining a variable proportional to the canonical momentum of the field $V_i^{\bot}$

\be
\zeta=-{V_e(\l) \, f\,e^{\Awf} \over   \omega} { \partial_r \psi_V \over \psi_V} \, .
\ee
The new variable satisfies the following first order equation

\be
\zeta'-{\omega \over f} \left({\zeta^2 \over V_e(\l) \, e^{\Awf}} + V_e(\l) 2\,  e^{\Awf} \left( 1-f {k^2 \over \omega^2} \right) \right)=0 \, .
\label{1ordeq}
\ee
The incoming regularity condition on the horizon for $\psi_V$ translates to $\zeta'(r_h)=0$, hence

\be
\zeta_h = i\, V_e(\l_h) \,e^{\Awf_h} \, .
\label{bchor}
\ee
where the functions with a subindex $h$ correspond to their values at the horizon $r=r_h$. The spectral density in terms of $\zeta$ is

\be
{ \rho(\omega) \over \omega}=6 M^{3} N_c N_f  \mathrm{Im }\zeta(\epsilon)\,.
\ee
In the numerical computation of $\rho$, we use the first order equation, (\ref{1ordeq}), since the numerical errors close to the boundary are reduced significantly,  \cite{Iatrakis:2014txa}.
The electric conductivity can be extracted from the time-like limit $\omega, k\rightarrow 0$ limit of the transverse
vector spectral function as 
\be
\sigma = \lim_{\omega \to 0} {\rho (\omega) \over 6 \, \omega}\;.
\ee
Solving \eqref{1ordeq} for ${\bf k}=0$ and $\omega=0$, with the boundary condition \eqref{bchor}, we  find

\be
\sigma=
M^{3} N_c N_f  V_e(\l_h) \,e^{\Awf_h} ,
\label{trcoef}
\ee
Equivalently, the electric conductivity can be obtained from the light-like spectral function through 
\be
\sigma = \lim_{|{\bf k}|=\omega \to 0} {\rho (\omega, {\bf k}) \over 4 \, \omega}\;.
 \ee
Therefore, the electric conductivity is proportional to photon-emission rate in low energy.

\subsection{Electric Conductivity and Photon-Emission Rates}

\begin{figure}[t]
	\begin{center}
		{\includegraphics[width=8cm,height=6cm,clip]{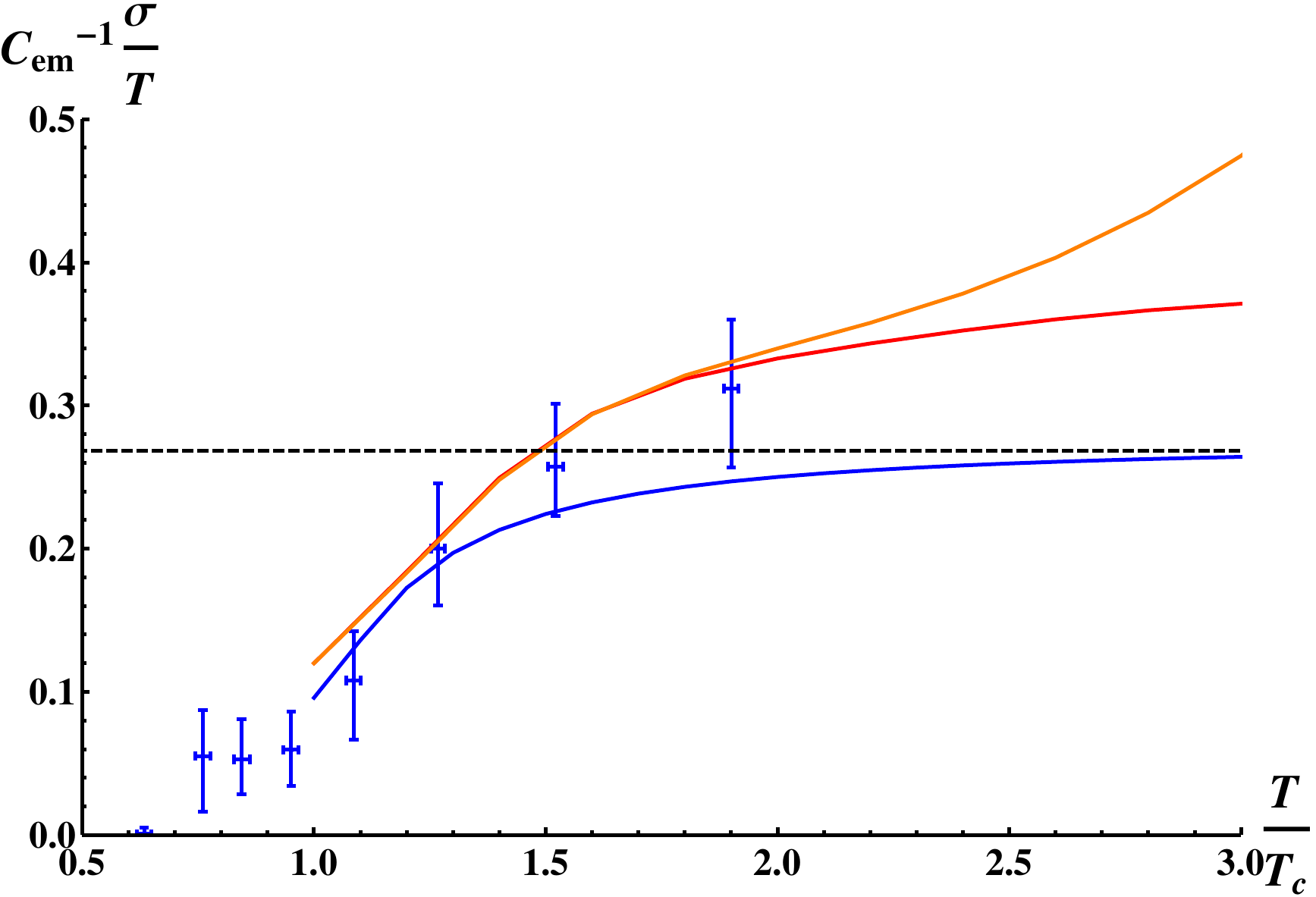}}
		\caption{The orange, red, and blue (from top to bottom ) curves correspond to VQCD2, VQCD1, and GN model, respectively. The black dashed line represents the result for the SYM plasma and the blue dots correspond to the lattice simulation with three flavors \cite{AartsAlltonAmatoEtAl2015}, where $C_{em}=2e^2/3$ and we take $N_c=N_f=3$.}\label{DC_conductivity}
	\end{center}
\end{figure}

%=======================================
\begin{figure*}[ht!]
	\centering
	\begin{tabular}{cc}
		\includegraphics[width=0.5\linewidth]{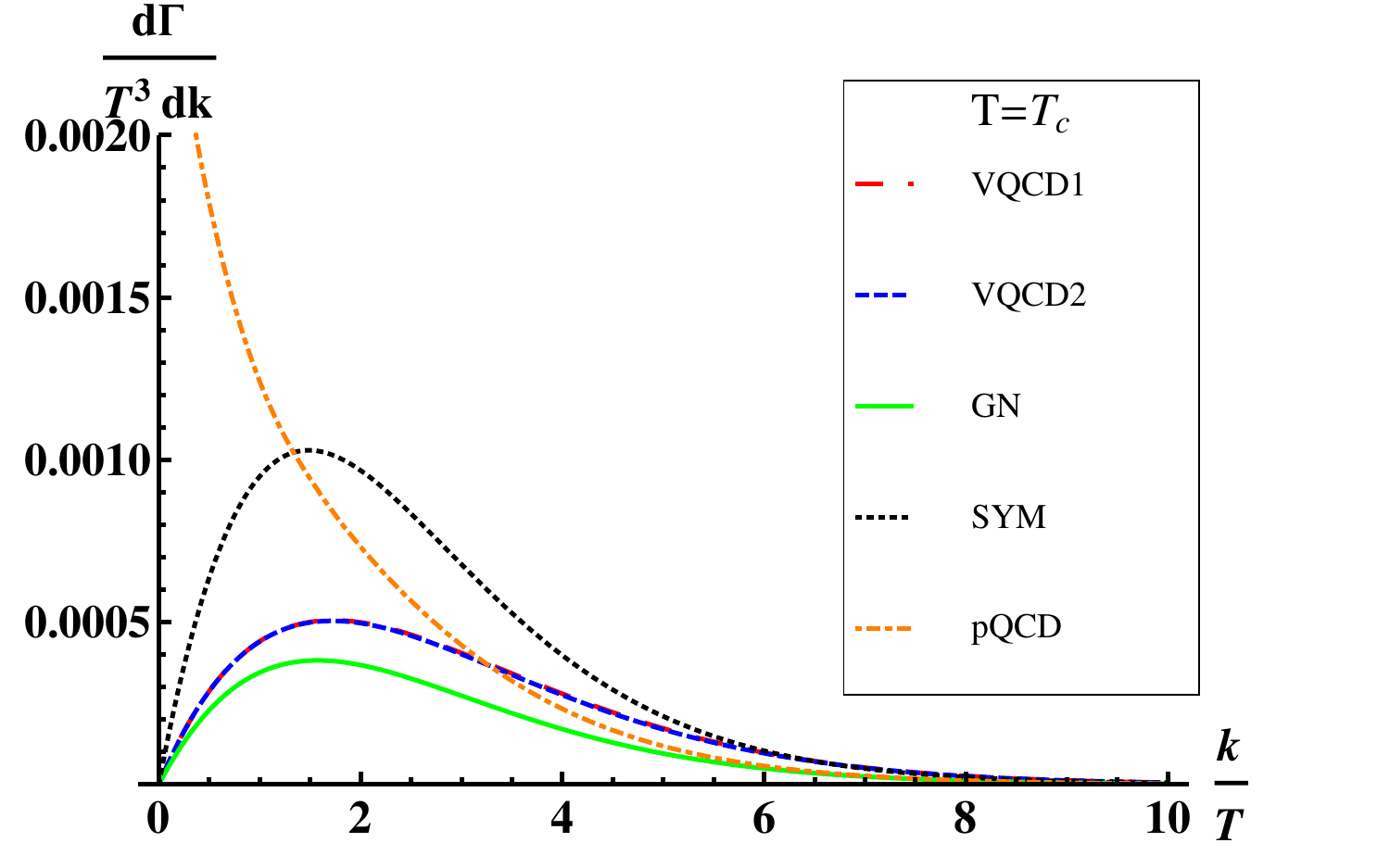} &
		\includegraphics[width=0.5\linewidth]{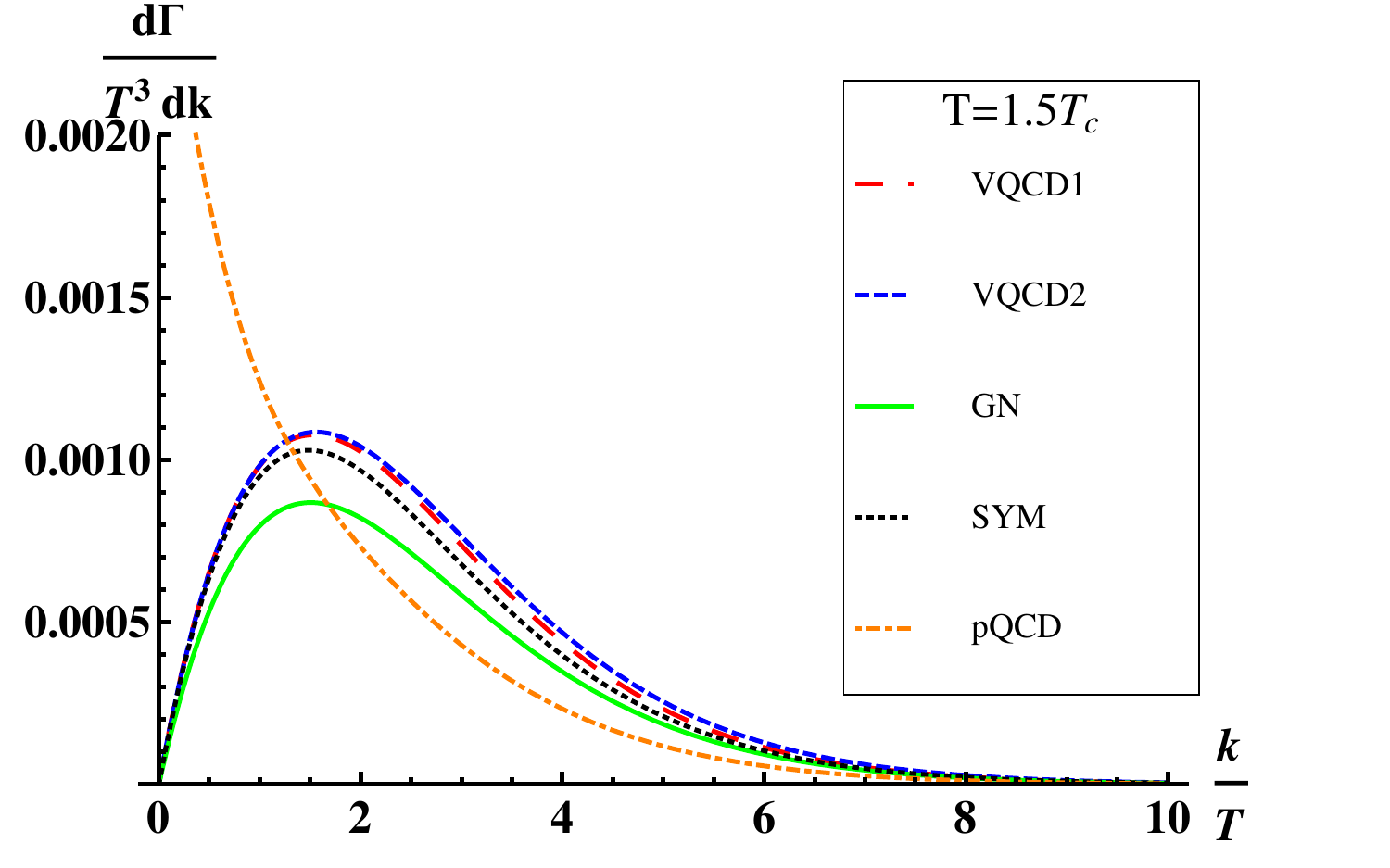} \\
		\includegraphics[width=0.5\linewidth]{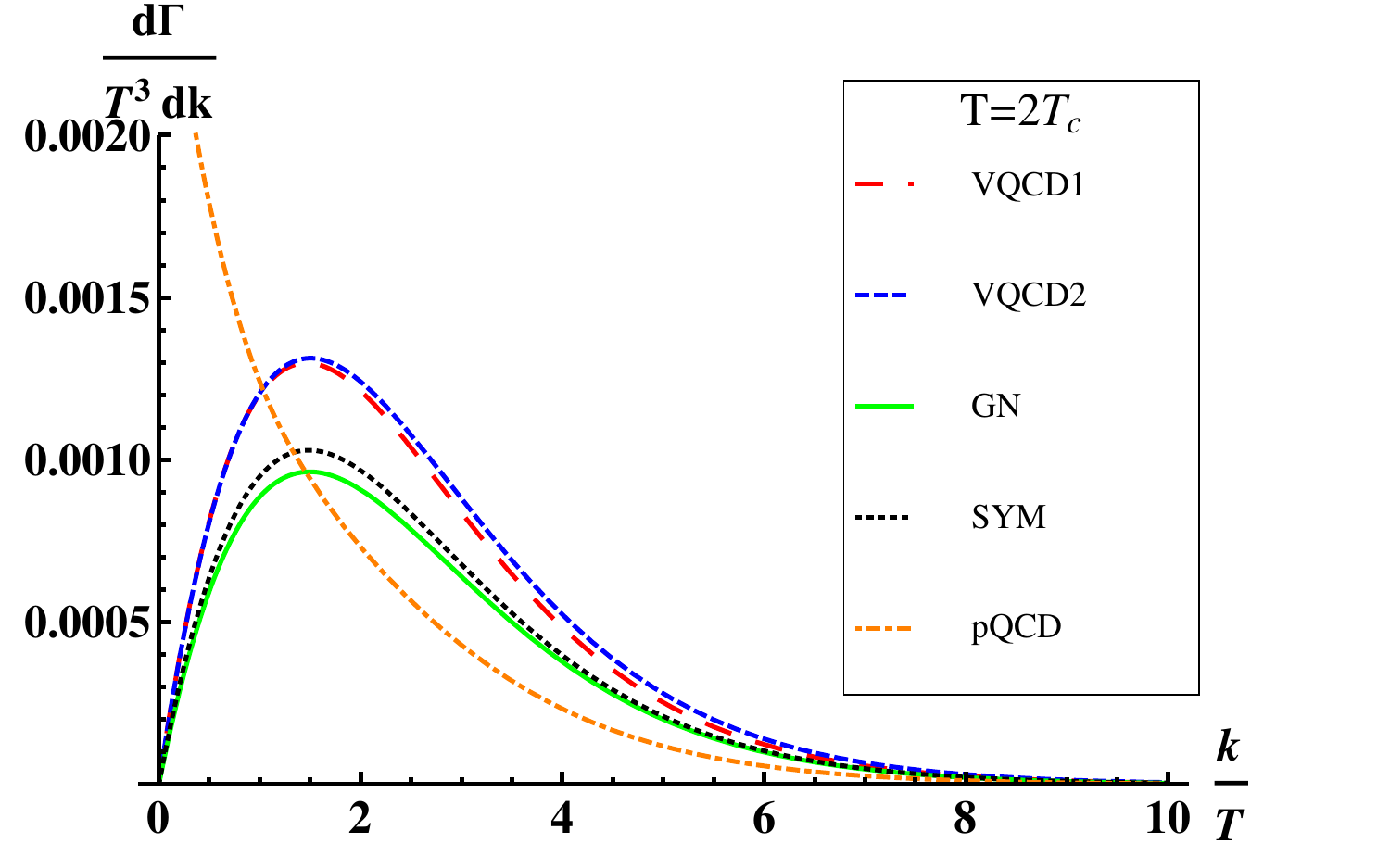} &
		\includegraphics[width=0.5\linewidth]{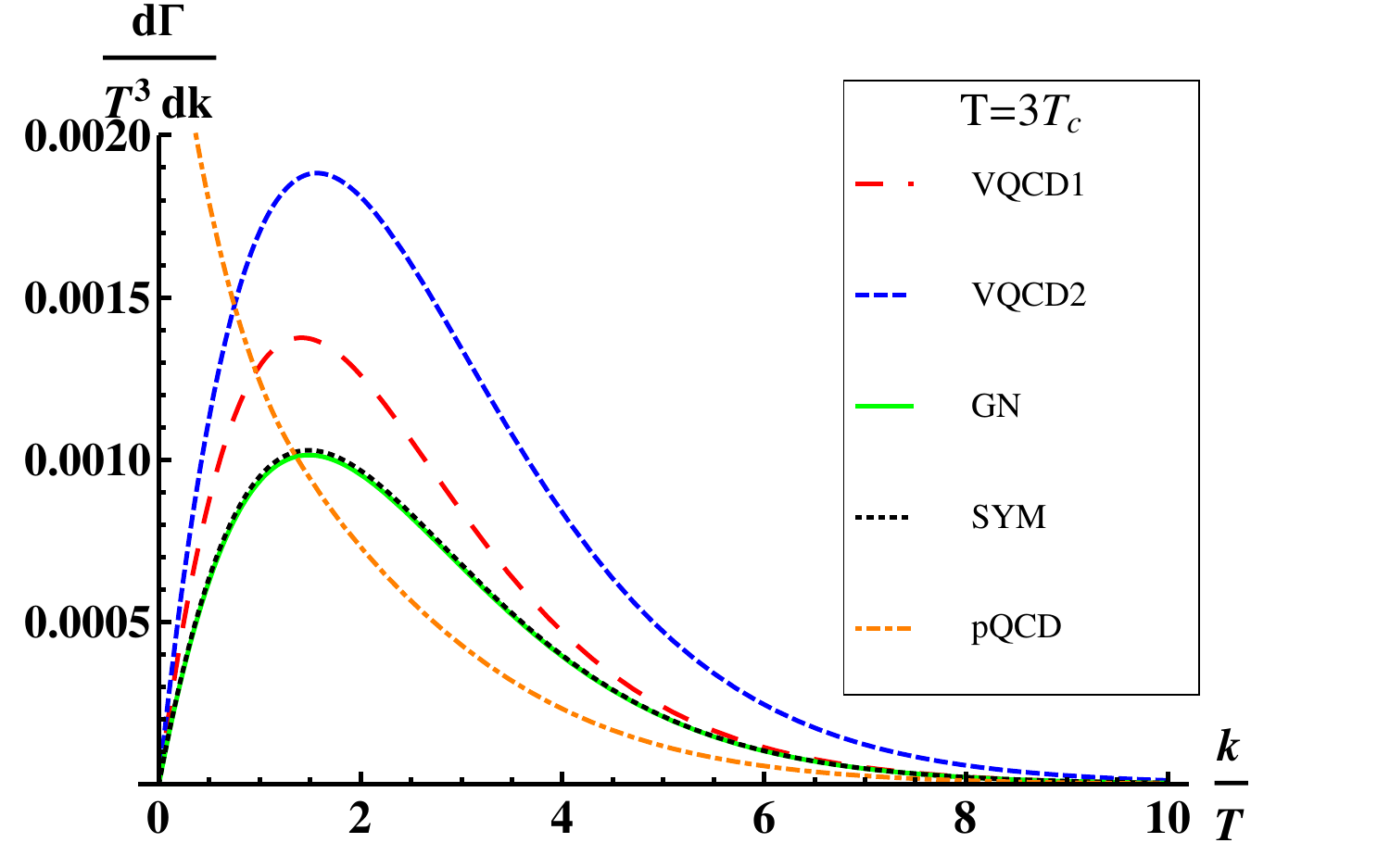}
	\end{tabular}
	\caption{Photon-emission rates from different models. The pQCD rates are for the weakly coupled QGP at $g_s=2$ at next-to-leading order. }
	\label{fig_photon_rates}
\end{figure*}
%=======================================

In Fig.\ref{DC_conductivity}, we present the electric conductivity from distinct models in comparison with the latest result from lattice simulations, where the extra normalization factor $C_{em}=2e^2/3$ comes from taking the trace over flavors. 

From holography, the conductivity rescaled by temperature for the strongly coupled SYM plasma has an analytic expression \cite{CaronHuot:2006te},
\begin{eqnarray}
\sigma/T=\frac{e^2N_c^2}{16\pi},
\end{eqnarray}
which is constant in $T$ due to conformal invariance. On the contrary, the most recent simulation for three-flavor QCD \cite{AartsAlltonAmatoEtAl2015} shows that the electric conductivity rescaled by $T$ monotonically decreases against $T$ near $T_c$ based on the conformal anomaly.
The GN model, which saturates the result for the SYM plasma at high $T$, roughly matches the lattice result around $T_c$ but starts to deviate for $T>1.5T_c$. On the other hand, the VQCD1 and VQCD2 approximately fit the lattice result above $T_c$. For $T>2T_c$, due to the lack of lattice results, we make different setup in the models such that VQCD2 leads to more rapid growth for $\sigma/T$.

In Fig.\ref{fig_photon_rates}, we present the photon-emission rates from different models compared with the ones from pQCD at the next-to- leading order based on hard-thermal-loop resummation \cite{Ghiglieri:2013gia}, %which includes the 2 to 2 scatterings \cite{Shen:2014nfa} and collinear emissions \cite{Arnold:2001ms}, 
where we take $N_c=N_f=3$ and $\alpha_{EM}=e^2/(4\pi)=1/137$.

In general, similar to the finding in the SYM plasma \cite{CaronHuot:2006te}, the emission rates from holography models have distinct features at low energy compared with the ones in the weakly coupled scenario. However, the divergence of pQCD rate in low energy stems from the breakdown of perturbative calculations for $k\lesssim 3T$ ($3T$ is the mean momentum in the plasma). On the other hand, the emission rate at small $k\sim g_s^4T$ softer than all other transport frequency scales in the weakly coupled plasma should be dictated by electric conductivity. Due to the shortage of a rigorous treatment of the emission rate at intermediate momenta, we employ phenomenological interpolation based on the relaxation-time approximation and linear extrapolation via the electric conductivity from lattice simulations to fix the IR divergence. The details of interpolation and corresponding results are shown in Appendix (see Ref.\cite{Ghiglieri:2016tvj} for a different approach of interpolation constrained by lattice simulations). In theses cases, the shapes of pQCD rates become more analogous to those of holographic models. Nonetheless, we find that the corrections of such IR behaviors for pQCD rates upon the spectra and anisotropic flow of QGP photons and direct photons are almost negligible.     

%However, it is worthwhile to note that the 2 to 2 collisions result in similar shapes of emission rates compared with those from holographic models, whereas the collinear emissions lead to IR divergence. In contrast, the photon production in holographic models incorporate both mechanisms, which "might" suggests that the collinear emissions are suppressed in strong coupling. 

As shown in the photon production from the SYM plasma with finite t'Hooft coupling in holography \cite{HassanainSchvellinger2012,Hassanain:2011ce}, the peak of the photon emission rate shifts to small $k/T$ when the coupling is reduced. The finding in holography seems to be consistent with the trend found in the weakly coupled scenario, where the emission rate has blueshift when the coupling is increased \cite{Arnold:2001ba,Arnold:2001ms,CaronHuot:2006te}. Nevertheless, unlike the SYM plasma and pQCD with a fixed coupling, non-conformal models now result in the increase of the rescaled rates with temperature, which is qualitatively in accordance with the results for $\sigma/T$. As shown in Fig.\ref{fig_photon_rates}, the increase is most prominent at $k/T\approx 2$. Although the positions of the peaks in $k/T$ approximately remain unchanged with $T$ for all holographic models and pQCD here, they should in fact shift to the infrared (IR) region in the unit of $k$ at lower temperature.

\section{Direct Photon Production in relativistic heavy-ion collisions}
\label{sec_spec_and_flow}

In this section, we explore the phenomenological significance of using different sets of the QGP photon emission rates in modeling the direct photon production in relativistic heavy-ion collisions.

\subsection{Model Setup}
In relativistic heavy-ion collisions there are many emission sources to produce photons \cite{Shen:2014thesis,Shen:2015nto,Shen:2016odt}. The dominant contribution to those photons produced with transverse momentum $p_T < 4$ GeV comes from prompt production and thermal radiation. In this work, we only consider these two sources for direct photons in the low $p_T$ region. Prompt photons include productions from the initial hard scattering processes (Compton scattering and $q\bar{q}$ annihilations) as well as from QCD jet fragmentation. The $p_T$-spectra of prompt photons in p-p collisions can be computed using perturbative QCD up to next-to-leading order (NLO) \cite{Aurenche:2006vj, Paquet:2015thesis}. 

The prompt photon production in nucleus-nucleus collisions is estimated using the number of binary collision $N_\mathrm{coll}$ scaled photon spectra in pp collisions with cold nuclear and isospin effects included \cite{Eskola:2009uj,Paquet:2015thesis}. Here, we used the parameterization presented in Ref.~\cite{Paquet:2015lta} for the prompt sources in Pb+Pb collisions at 2.76 $A$\,TeV and in Au+Au collisions at 200 $A$\,GeV.

To compute thermal photon production from expanding relativistic heavy-ion collisions, we convolute the thermal photon emission rates with event-by-event hydrodynamic medium. We use the state-of-the-art medium evolution as described in Ref.~\cite{Ryu:2015vwa,Paquet:2015lta}. The hydrodynamic simulations are tightly constrained by various hadronic observables.  Event-by-event IP-Glasma initial conditions are matched to hydrodynamics at $\tau_\mathrm{sw} = 0.4$ fm/c and then evolved with lattice QCD based equation of state, s95p-v1 \cite{Huovinen:2009yb,Shen:2010uy}. Both shear and bulk viscous effects are included in the medium evolution.

Thermal photon radiation is considered from fluid cells whose temperatures are above 105 MeV. The different sets of photon production rates in the QGP phase, as described in the previous section, are applied to the $T > T_c (= 180$\,MeV) temperature  region. For the QGP rates derived from a weakly-coupled QCD plasma, 2 to 2 scatterings \cite{Shen:2014nfa} and collinear emissions \cite{Arnold:2001ms} are included, which are referred as pQCD rates in previous sections. In the hadronic phase, whose temperature is  between 105 and 180 MeV, we use the current most complete set of hadronic photon emission rates, which includes contribution from meson-meson reactions \cite{Turbide:2003si}, many-body $\rho$-spectral function \cite{Rapp:1999ej,Rapp:1999qu}, $\pi-\pi$ bremsstrahlung \cite{Liu:2007zzw,Heffernan:2014mla}, and emission from $\pi-\rho-\omega$ interactions \cite{Holt:2015cda}. 

At the current stage, shear and bulk viscous corrections are only known for a subset of photon production channels. In order to make fair comparisons among the different sets of QGP rates, we will not include any viscous corrections to photon emission rates in this work. Detailed analysis of the viscous effects on direct photon observables were presented in Refs.~\cite{Shen:2013cca,Shen:2014cga,Shen:2014nfa,Paquet:2015lta}.

The momentum distribution of thermal photons is computed by first producing photons in the local rest of frame of every fluid cell,
whose temperature $T(x)$ is higher than the system's freeze-out temperature $T_\mathrm{freezeout}$. Then these photons are boosted with the corresponding fluid velocity $u(x)$
to the lab frame,
\begin{equation}
	q \frac{dN_\mathrm{thermal}^\gamma}{d^3 q} = \int_{T > T_\mathrm{freezeout}} d^4 x \bigg[ q \frac{dR^\gamma}{d^3 q}\left(T(x), E_q \right)\bigg\vert_{E_q = q \cdot u(x)} \bigg],
\end{equation}
%\textcolor{red}{Should we define all the symbols in the above formula?}
where the thermal photon emission rate is denoted as $q \frac{dR^\gamma}{d^3 q}$. The direct photon spectra is the sum of thermal and prompt photons,
\begin{equation}
q \frac{dN_\mathrm{direct}^\gamma}{d^3 q} = q \frac{dN_\mathrm{thermal}^\gamma}{d^3 q} + q \frac{dN_\mathrm{prompt}^\gamma}{d^3 q}.
\end{equation}
Because the thermal photon production rate is suppressed by $\sim\alpha_\mathrm{EM}/\alpha_S$ compared to hadrons, they are rare in heavy-ion collisions. Thus, the direct photons anisotropic flow coefficients need to be evaluated by correlating with the majority of the soft charged hadrons to ensure sufficient statistics.
This is known as the scalar-product method \cite{Luzum:2012da}. In the theoretical calculations,
the scalar-product direct photon anisotropic flow coefficients, $v^\gamma_n\{\mathrm{SP}\}$,
can be computed as \cite{Shen:2014thesis, Shen:2014lpa},
\begin{equation}
	v^\gamma_n\{\mathrm{SP}\}(p_T) = \frac{\langle v^\gamma_n(p_T) v^\mathrm{ref}_n \cos[n(\Psi_n^\gamma(p_T) - \Psi^\mathrm{ref}_n] \rangle}{\sqrt{\langle v_n^\mathrm{ref} \rangle^2}},
\end{equation}
where $v^\gamma_n(p_T)$ and $\Psi_n^\gamma(p_T)$ are the magnitude and angle of the $n$-order harmonic flow for direct photons at a given transverse momentum $p_T$. They are defined as the Fourier coefficients of the photon momentum distribution,
\begin{equation}
v_n^\gamma(p_T) e^{i n \Psi_n^\gamma(p_T)} = \frac{\int d \phi \frac{dN^{\gamma}}{dy p_T dp_T d\phi} e^{in\phi}}{\int d \phi \frac{dN^{\gamma}}{dy p_T dp_T d\phi}}.
\label{eq39}
\end{equation}
The reference flow $v^\mathrm{ref}_n$ and its flow angle $\Psi^\mathrm{ref}_n$ are computed using soft charged hadrons with momentum between 0.2 to 3.0 GeV. The definitions of $v^\mathrm{ref}_n$ and $\Psi^\mathrm{ref}_n$ are similar to (\ref{eq39}).
%\textcolor{red}{How are the angles defined?}

\subsection{Direct photon spectra in relativistic heavy-ion collisions}

%=======================================
\begin{figure*}[ht!]
	\centering
	\begin{tabular}{cc}
		 \includegraphics[width=0.48\linewidth,height=0.58\linewidth]{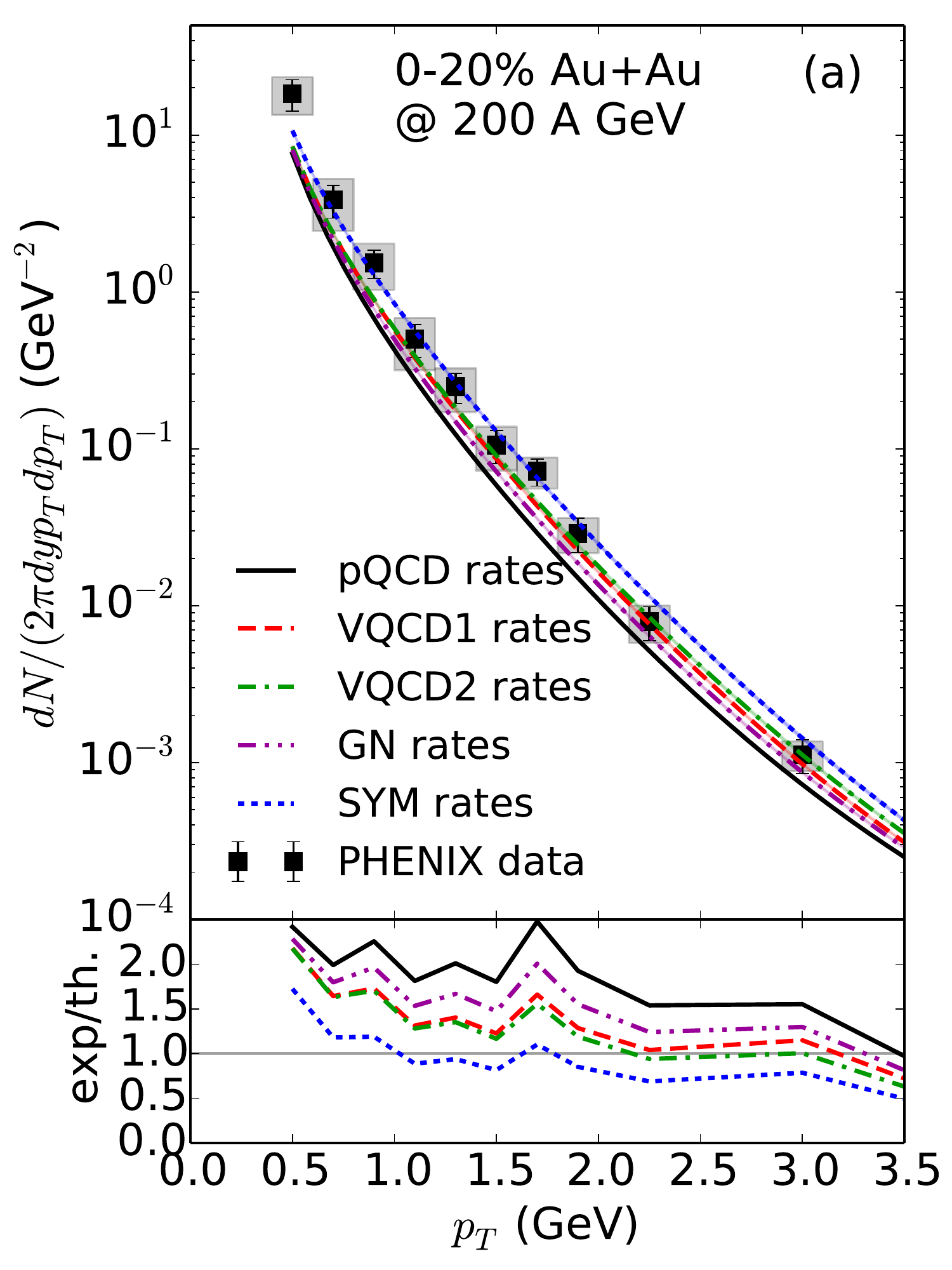} &
		 \includegraphics[width=0.48\linewidth,height=0.58\linewidth]{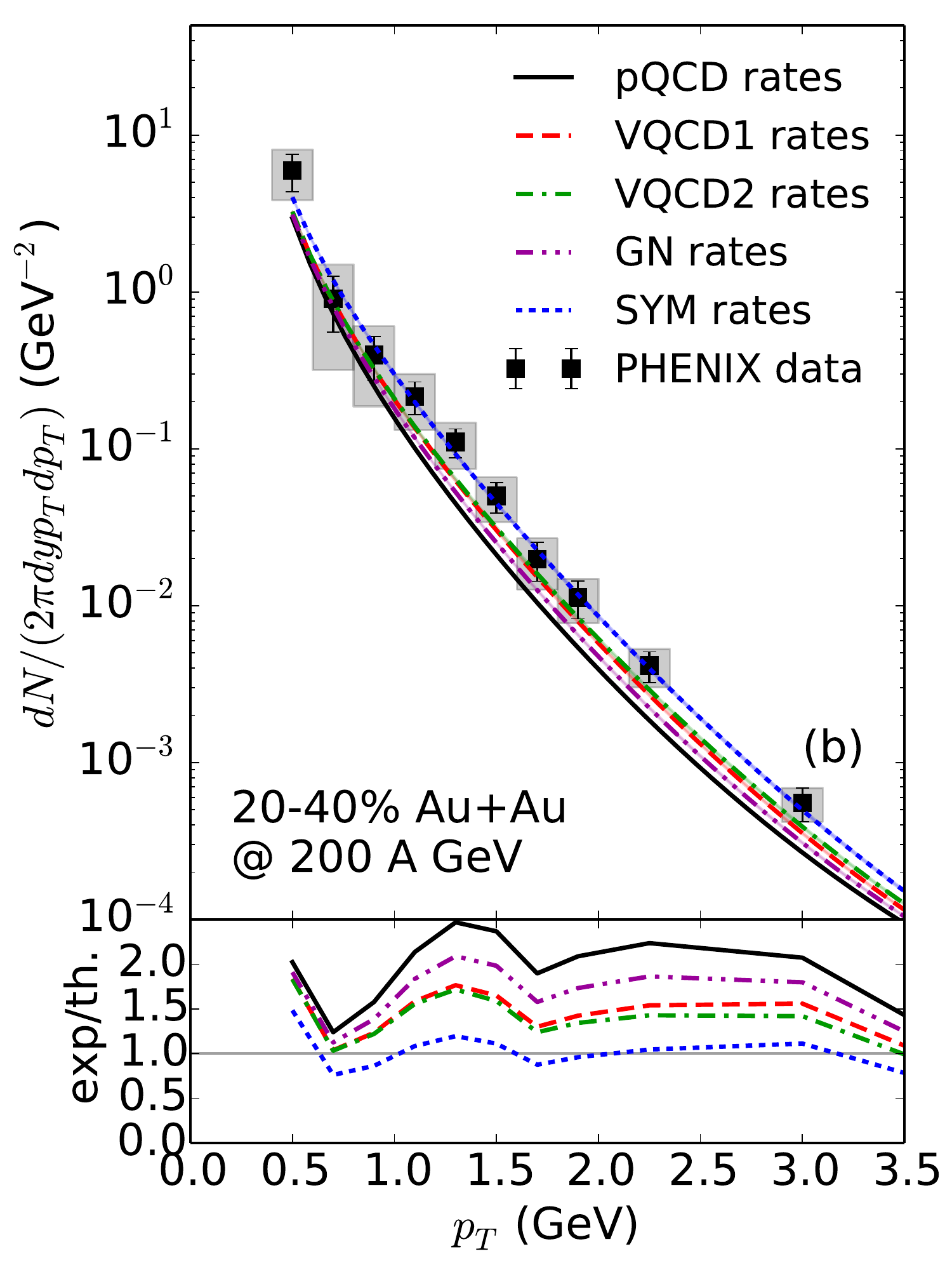} \\
		 \includegraphics[width=0.48\linewidth,height=0.58\linewidth]{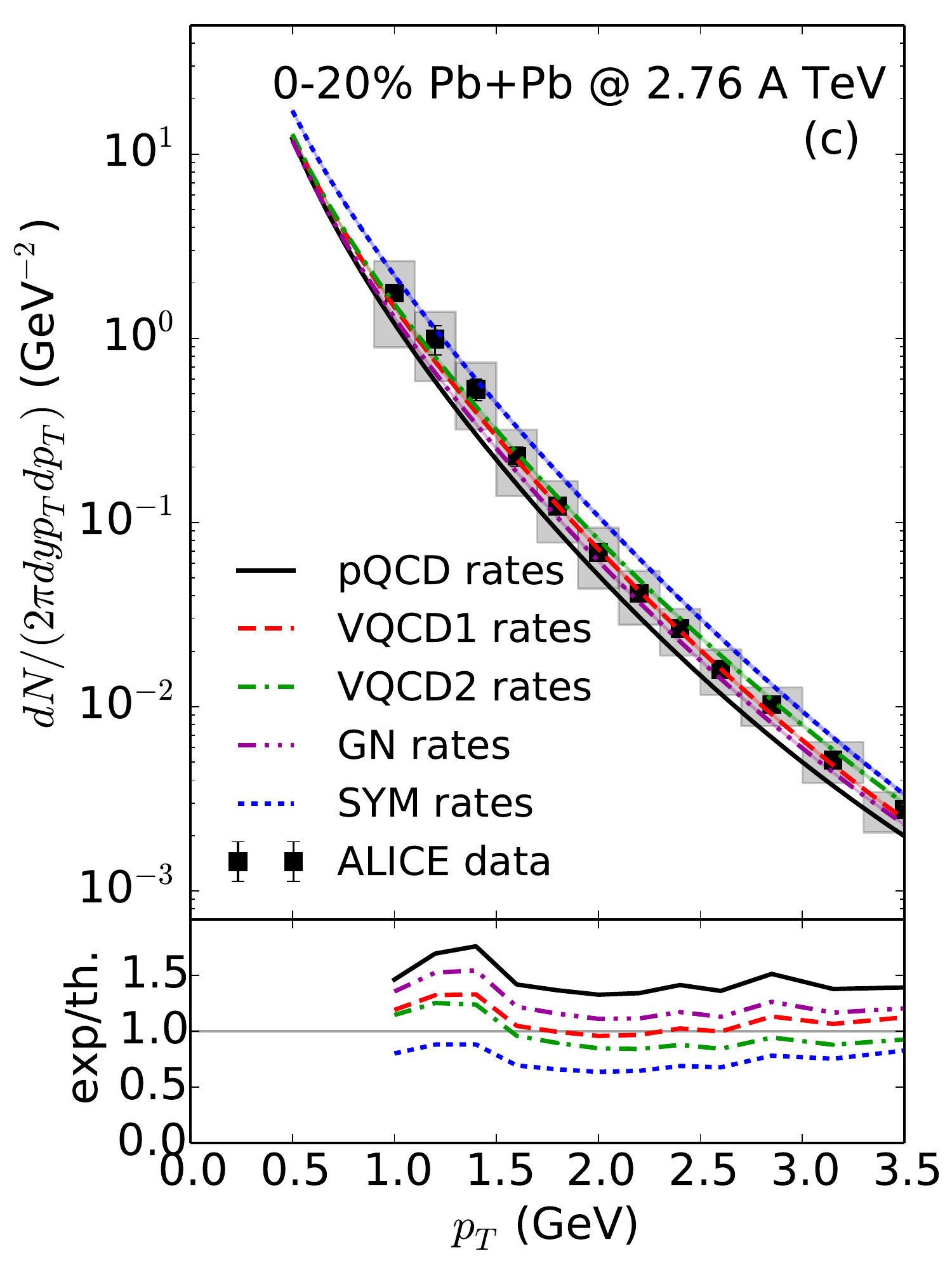} &
		 \includegraphics[width=0.48\linewidth,height=0.58\linewidth]{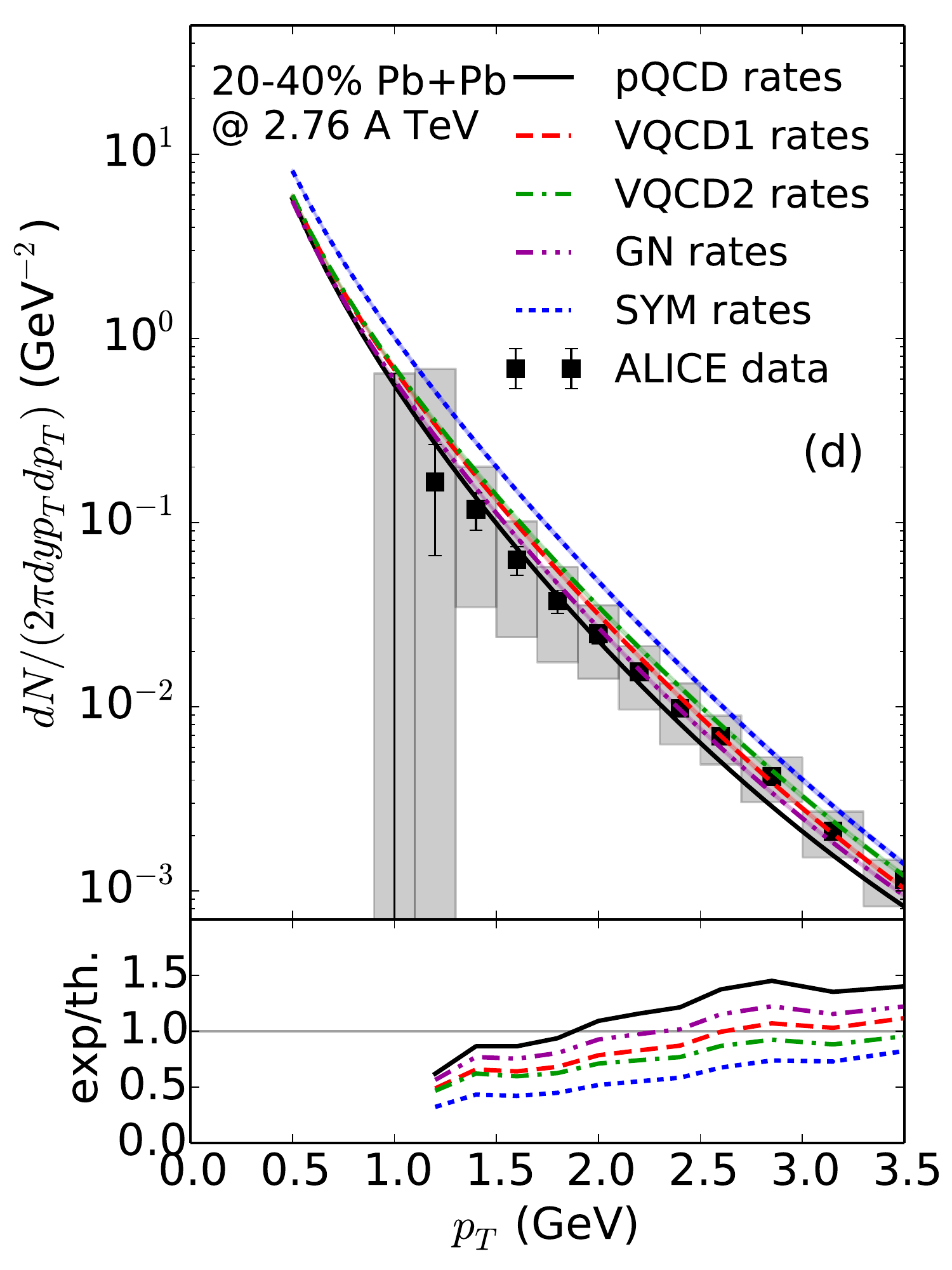}
	\end{tabular}
	\caption{(Color online) Direct photon spectra from 0-20\% (a) and 20-40\% (b) Au+Au collisions at 200 GeV compared with the PHENIX measurements \cite{Adare:2014fwh} and from 0-20\% (c) and 20-40\% (d) Pb+Pb collisions at 2.76 $A$\,TeV compared with the ALICE measurements \cite{Adam:2015lda}. The ratios of experimental data to theoretical results are shown in the bottom of each plot.}
	\label{fig6.1}
\end{figure*}
%=======================================
%
Direct photon spectra using different sets of QGP photon rates are compared to the experimental measurements in Au+Au collisions at 200 $A$\,GeV at the RHIC and in Pb+Pb collisions at 2.76 $A$\,TeV at the LHC in Figs.~\ref{fig6.1}.
The QGP photon rates from holographic models result in more thermal radiation compared to the results with the pQCD rate. 
The reasons for this depend on the holographic model.
First, at strong coupling we expect more photon emissions than at weak coupling. On top of this, SYM which contains extra supersymmetric partners is expected to give the highest rate and this is turns out to be correct. The GN and VQCD models are non supersymmetric and have the same number of (perturbative) degrees of freedom as QCD.   

Therefore, among the different holographic rates, the SYM rates give the most direct photons.
Although the electric conductivities in the two VQCD models exceed the one in SYM model for $T > 1.5 T_c$, the direct photon yields from the VQCD models are smaller compared to the SYM results. This is because that most of the thermal photon radiations are coming from the phase transition region, $150 < T < 250$ MeV \cite{Shen:2013cca}, where the space-time volume is the largest. In this temperature region, the photon emission rates are suppressed in the VQCD models compared to the SYM rates. On the other hand, the GN model leads to smaller spectra compared to the ones for VQCD and SYM models as expected from the electric conductivity and emission rates.

%=======================================
\begin{figure*}[ht!]
	\centering
	\begin{tabular}{cc}
		\includegraphics[width=0.48\linewidth]{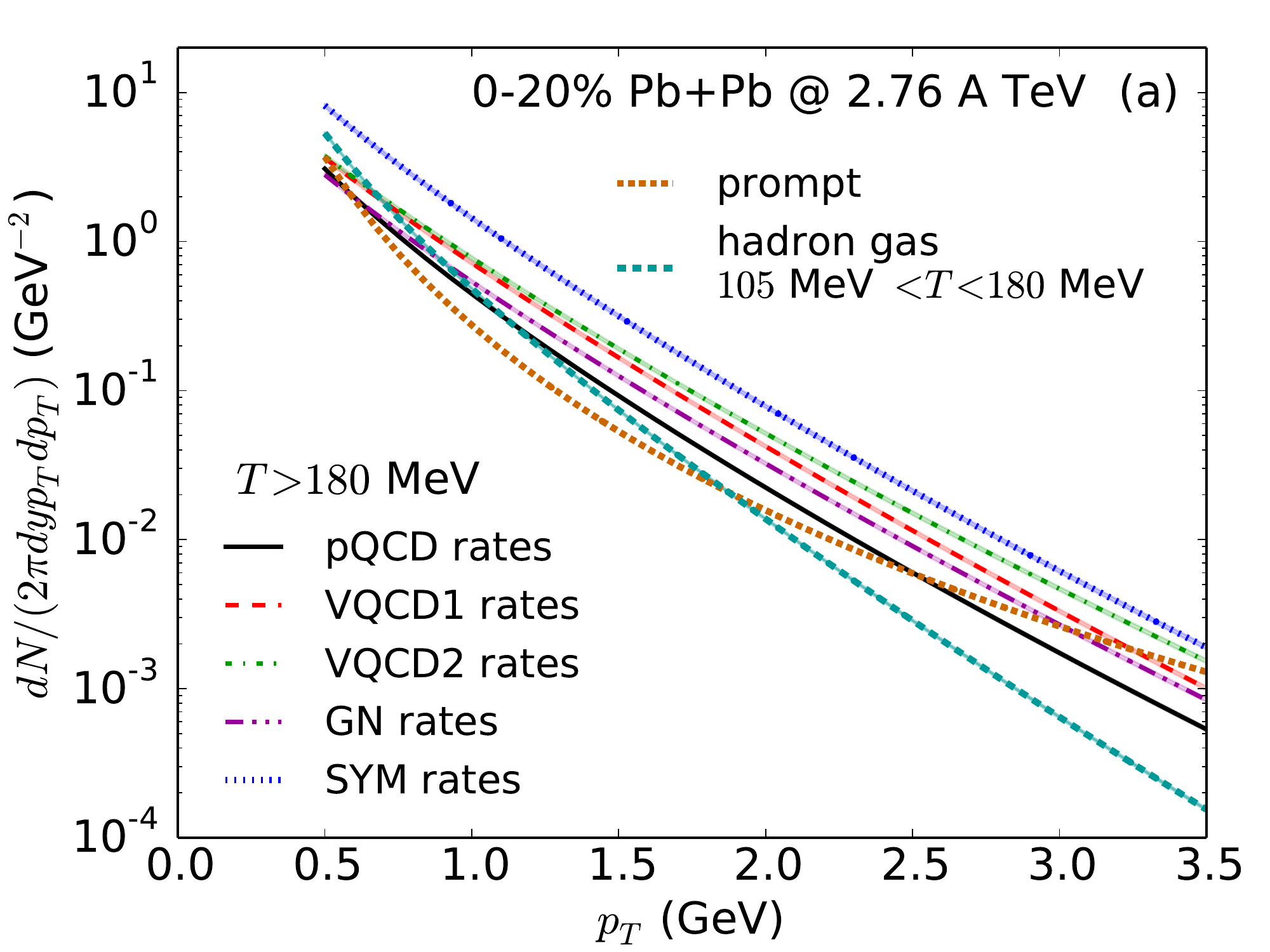} &
		 \includegraphics[width=0.48\linewidth]{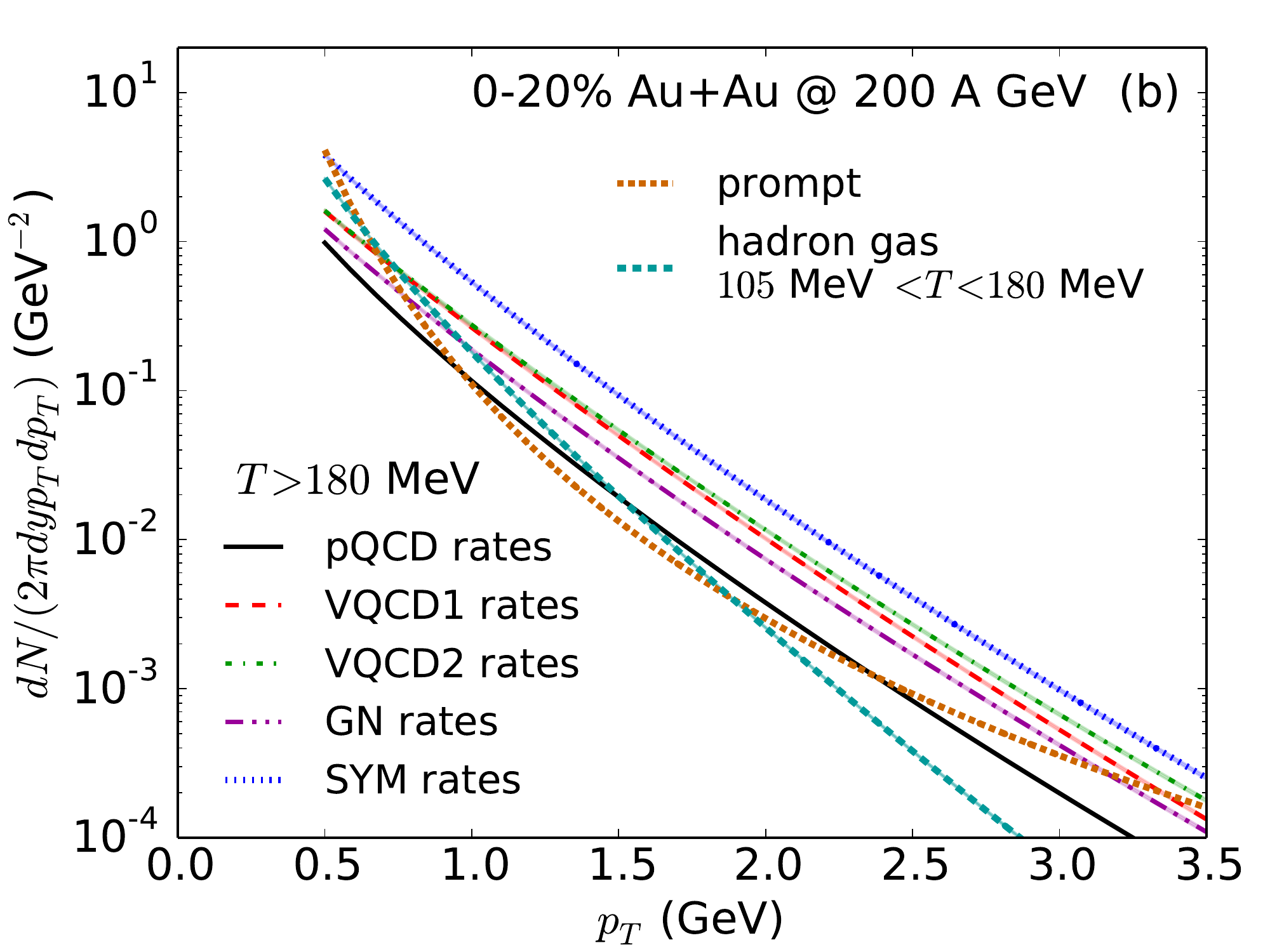} \\
	\end{tabular}
	\caption{(Color online) Comparisons of thermal QGP photon spectra using different sets of photon emission rates from $T > 180$ MeV regions in 0-20\% Pb+Pb collisions at 2.76 A TeV $(a)$ and 0-20\% Au+Au collisions at 200 GeV $(b)$. Spectra of hadronic photons and prompt photons are also shown.}
	\label{fig6.1.2}
\end{figure*}
%=======================================
%

In Fig.~\ref{fig6.1.2}, individual contribution to thermal photons are directly compared among different sets of QGP photon emission rates. Compared to prompt and hadronic radiation components, thermal photons from the QGP phase are dominant for $1.0 < p_T < 2.5$ GeV in both 0-20\% Pb+Pb collisions at 2.76 $A$\,TeV and 0-20\% Au+Au collisions at 200 $A$\,GeV.

This $p_T$ window which is dominated by the QGP photon emission is slightly larger with the holographic rates. The thermal radiation with the SYM rates is about factor of 2 larger than the results using other rates. The VQCD2 rates give the flattest slope of the thermal photon spectra, i.e. largest mean $p_T$. This is because there are more photons radiated from high temperature fluid cells by construction, as for $\l$ large we took $w(\l)$ to be larger than in other models.
The thermal radiation from the SYM rates is about factor of 2 larger than the results using other rates.

As the collision energy increases from RHIC to LHC, both the thermal radiation and prompt photon productions increase. The growth of the prompt component is a little bit faster than the thermal radiation. A similar trend was found in the predictions at even higher collision energy~\cite{Chang:2015hqa}.

\subsection{Direct photon anisotropic flow coefficients }

On the one hand, the absolute yield of direct photon spectra provides  information about the system's space-time volume as well as the degrees of freedom of photon emitters in the medium. On the other hand, the anisotropic flows of direct photons are more sensitive to the relative temperature dependence of photon rates and their interplay with the development of hydrodynamic anisotropic flows during the evolution.

%=======================================
\begin{figure*}[ht!]
	\centering
	\begin{tabular}{cc}
		 \includegraphics[width=0.48\linewidth]{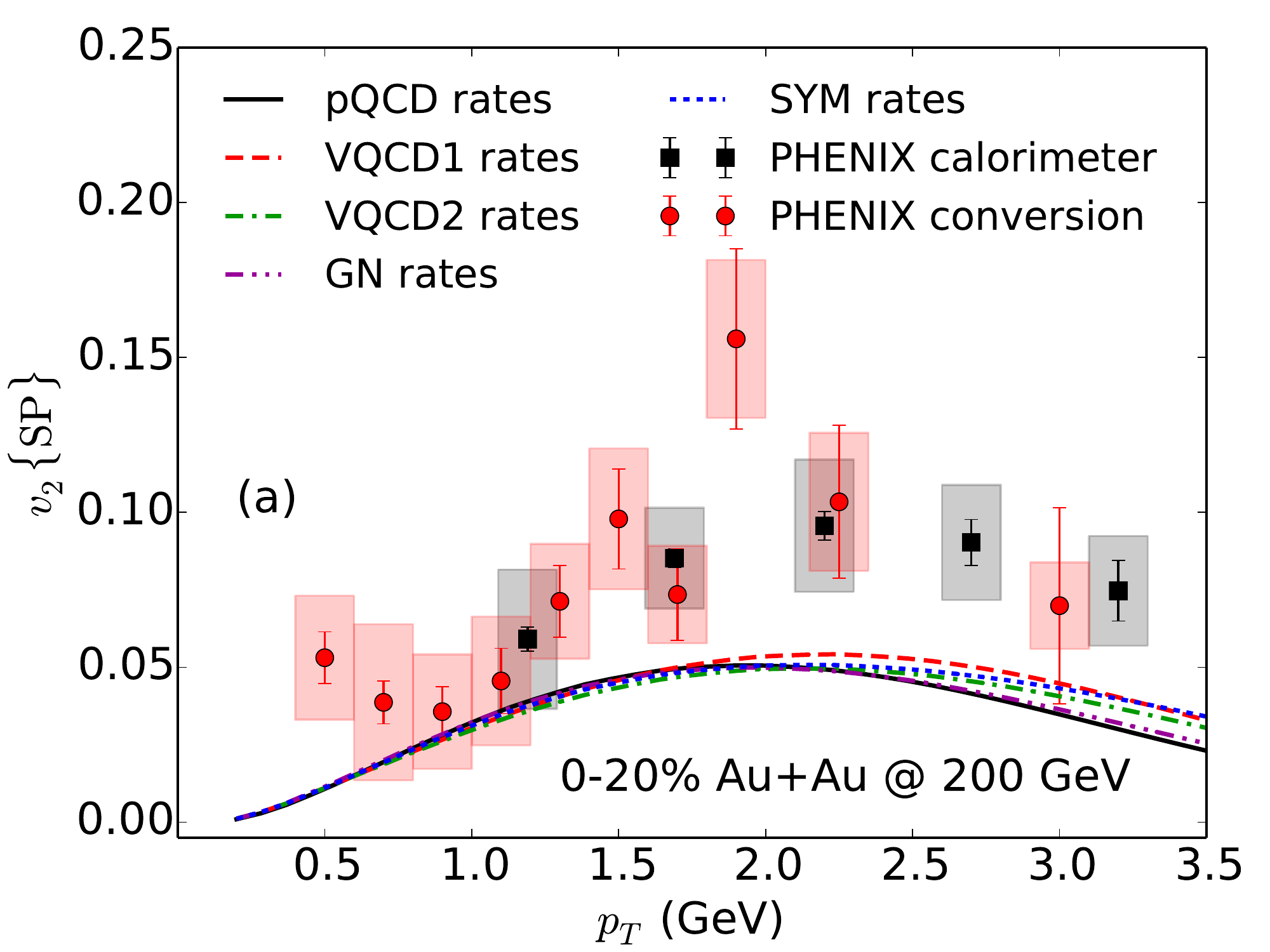} &
		 \includegraphics[width=0.48\linewidth]{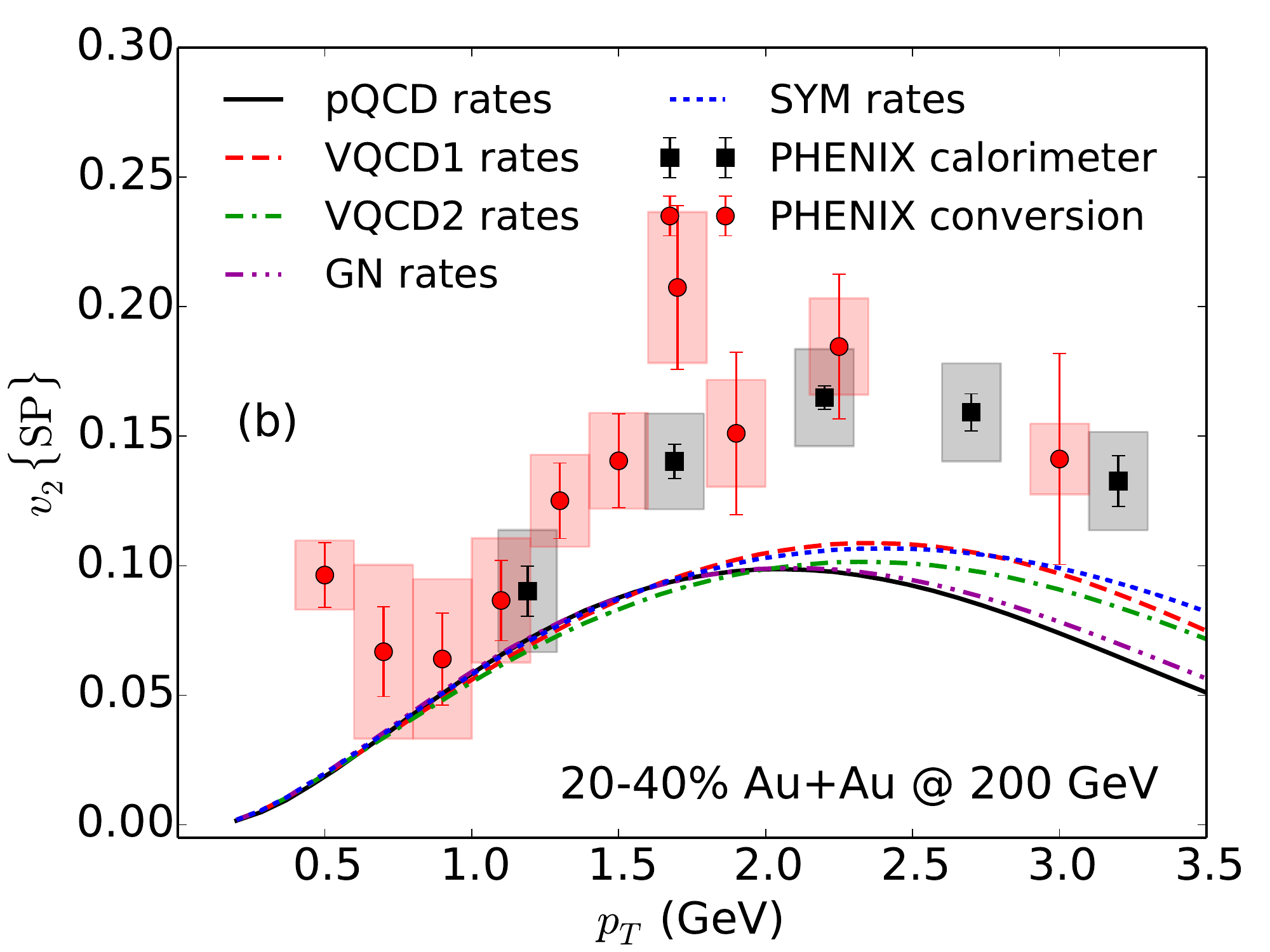} \\
		 \includegraphics[width=0.48\linewidth]{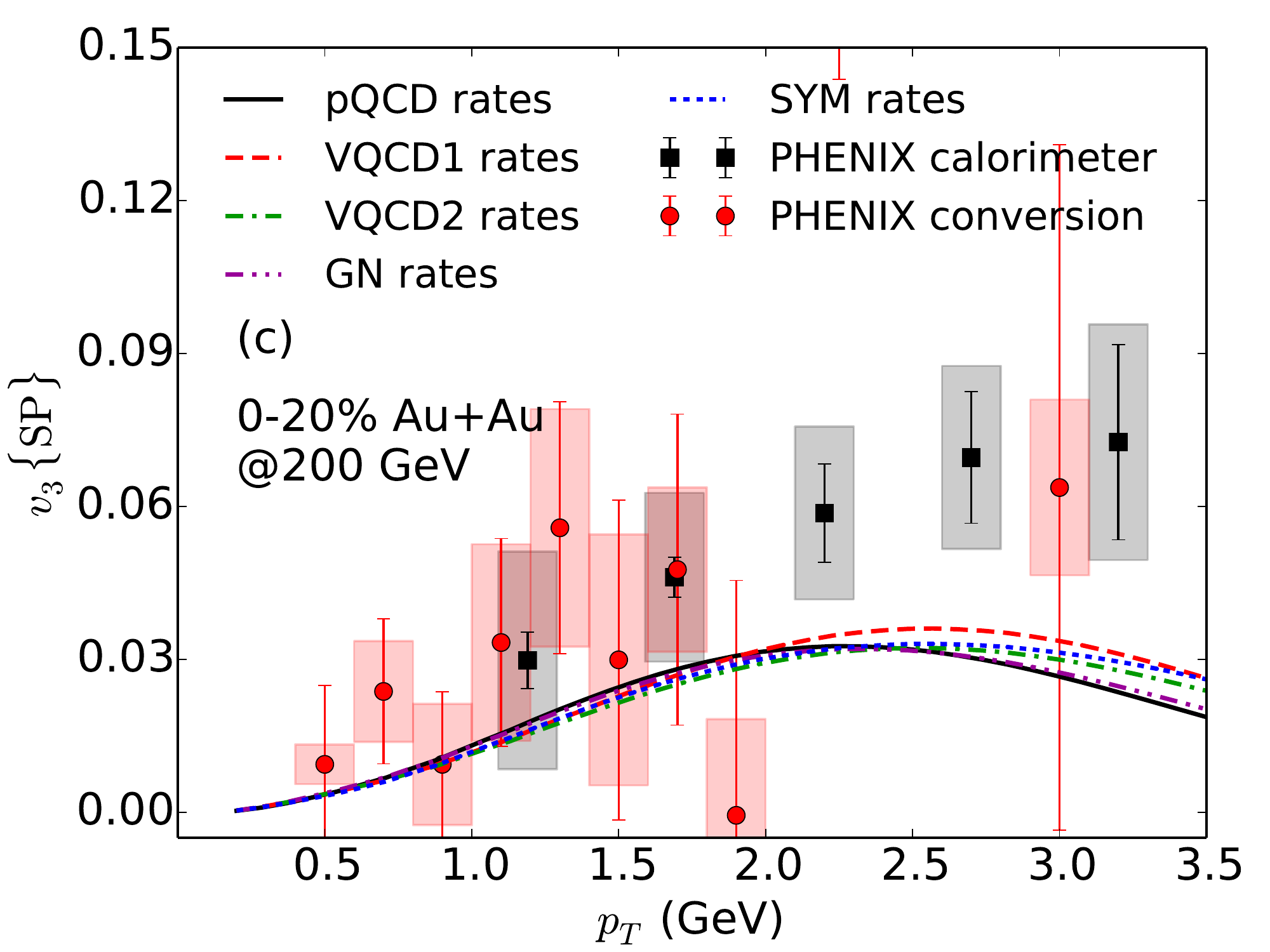} &
		 \includegraphics[width=0.48\linewidth]{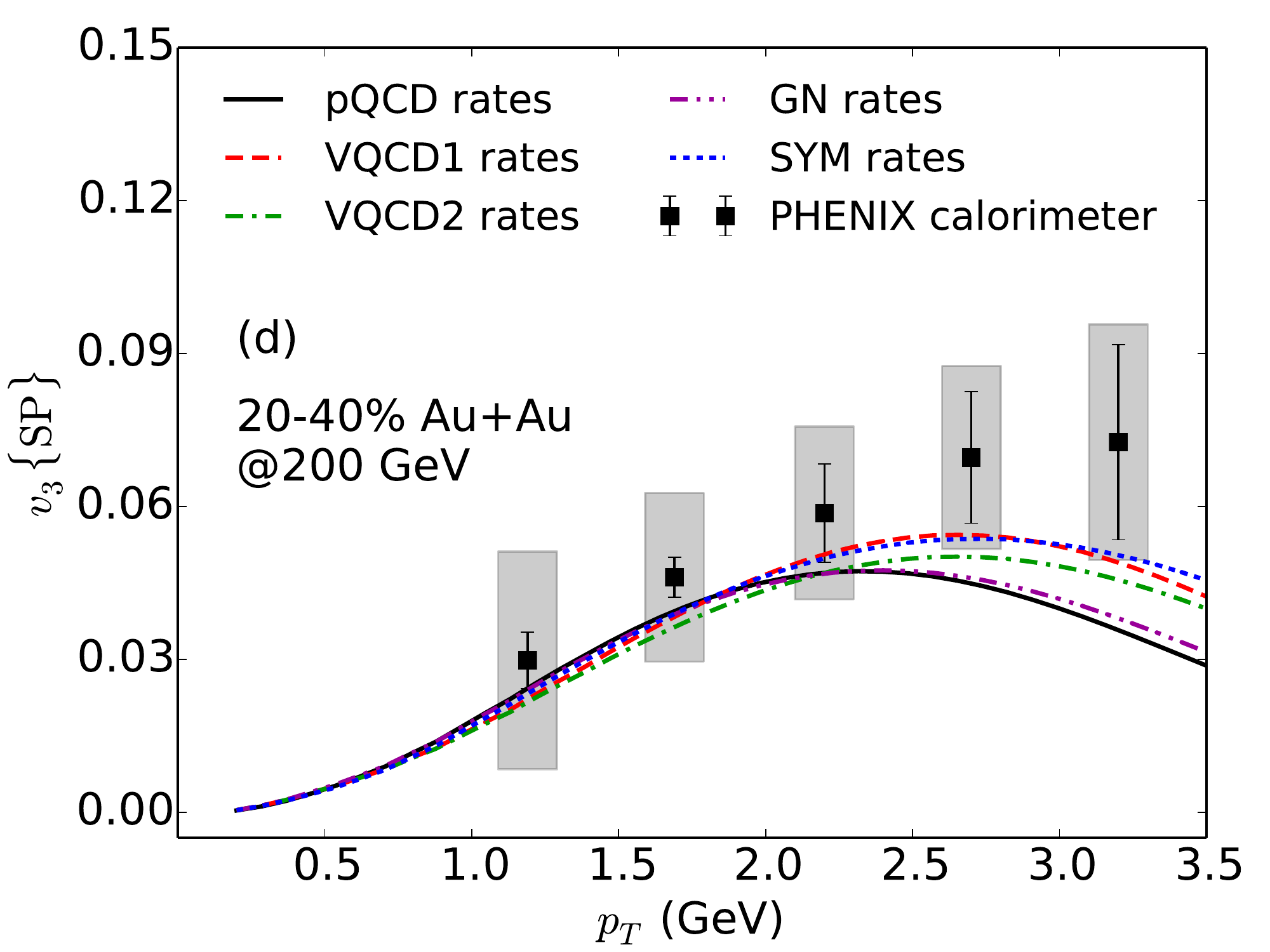}
	\end{tabular}
	\caption{(Color online) Direct photon anisotropic flow $v_{2,3}$ from 0-20\% (a,c) and 20-40\% (b,d) Au+Au collisions at 200 GeV compared with the PHENIX measurements \cite{Adare:2015lcd}.  }
	\label{fig6.2}
\end{figure*}
%=======================================
%
%=======================================
\begin{figure}[ht!]
	\centering
	\includegraphics[width=0.5\linewidth]{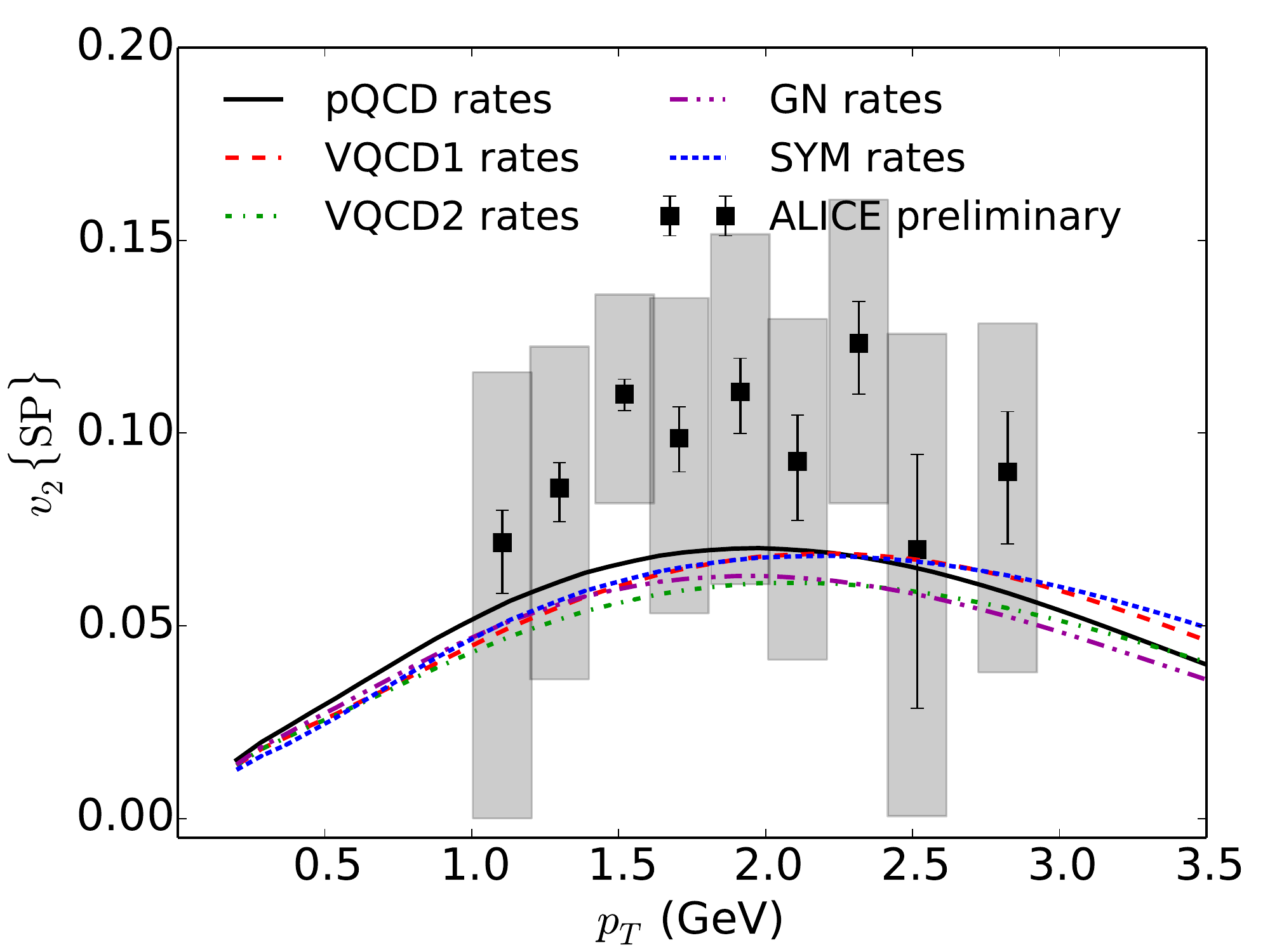}
	\caption{(Color online) Direct photon anisotropic flow $v_{2}$ in 0-40\% Pb+Pb collisions at 2.76 $A$\,TeV compared with the ALICE measurements \cite{Lohner:2012ct,Lohner:2013thesis}. }
	\label{fig6.3}
\end{figure}
%=======================================
%

In Figs.~\ref{fig6.2} and \ref{fig6.3}, direct photon anisotropic flow coefficients, $v_{2,3}\{\mathrm{SP}\}(p_T)$, are shown at the RHIC and LHC energies together with the experimental measurements. Since the underlying hydrodynamic medium is kept fixed for all sets of calculations, we here show curves without statistical error bands for better visual comparisons. At both collision energies, the weakly-coupled QCD rates gives the largest direct photon $v_n$ in the intermediate $p_T$ region, $1 < p_T < 2.5$ GeV. At the higher $p_T > 3.0$ GeV, the $v_n$ results using holographic rates are larger. This interesting hierarchy of direct photon $v_{2,3}$ is a results of the interplay between the temperature dependence of emission rates and the space-time structure of the hydrodynamic flow distribution.

\subsection{Direct photon emission in small collision systems}

Recently, sizable thermal radiation was found in high multiplicity light-heavy collisions in Ref.~\cite{Shen:2015qba}. Owing to large pressure gradients, small collision systems, such as p+Pb and d+Au collisions, expand more rapidly compared to the larger Au+Au collisions. This leads to a smaller hadronic phase in these collision systems. Most of the thermal photons come from the hot QGP phase, $T > 180$ MeV \cite{Shen:2015qba}. Hence, the difference between the QGP photon emission rates should be more distinctive in these small collision systems.

%\EK{Please read the two paragraphs below that I had to restructure to see if you agree.}
%\textcolor{blue}{DY: I agree.} 
In contrast to nucleus-nucleus collisions that were analyzed  in the previous sections, full (3+1)D hydrodynamic simulations with Monte-Carlo Glauber model as initial conditions are employed for the medium evolution in central p+Pb and d+Au collisions \cite{Shen:2016zpp}. The parameters in the hydrodynamic model are chosen such that a variety of hadronic flow observables can be reproduced. Starting at an initial proper time $\tau_0 = 0.6$ fm/$c$, every fluctuating energy density profile is evolved in full 3+1 dimensions with $\eta/s = 0.08$ for d+Au collisions or with $\eta/s = 0.10$ for p+Pb collisions.

In this work, we will use these calibrated hydrodynamic medium to study the sensitivity of direct photon observables in small systems to the different sets of QGP photon emission rates.

%=======================================
\begin{figure*}[ht!]
	\centering
	\begin{tabular}{cc}
		 \includegraphics[width=0.48\linewidth]{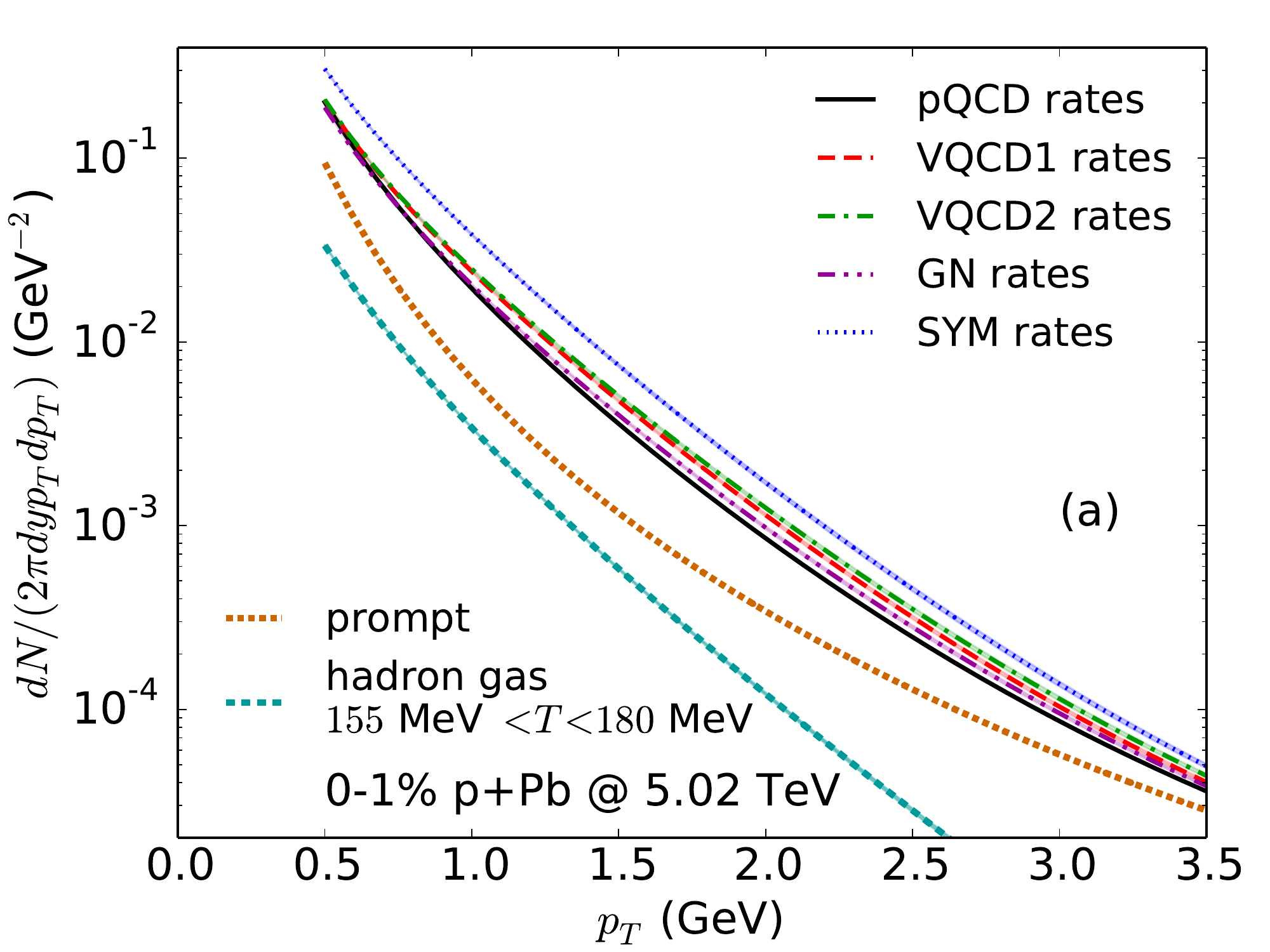} &
		 \includegraphics[width=0.48\linewidth]{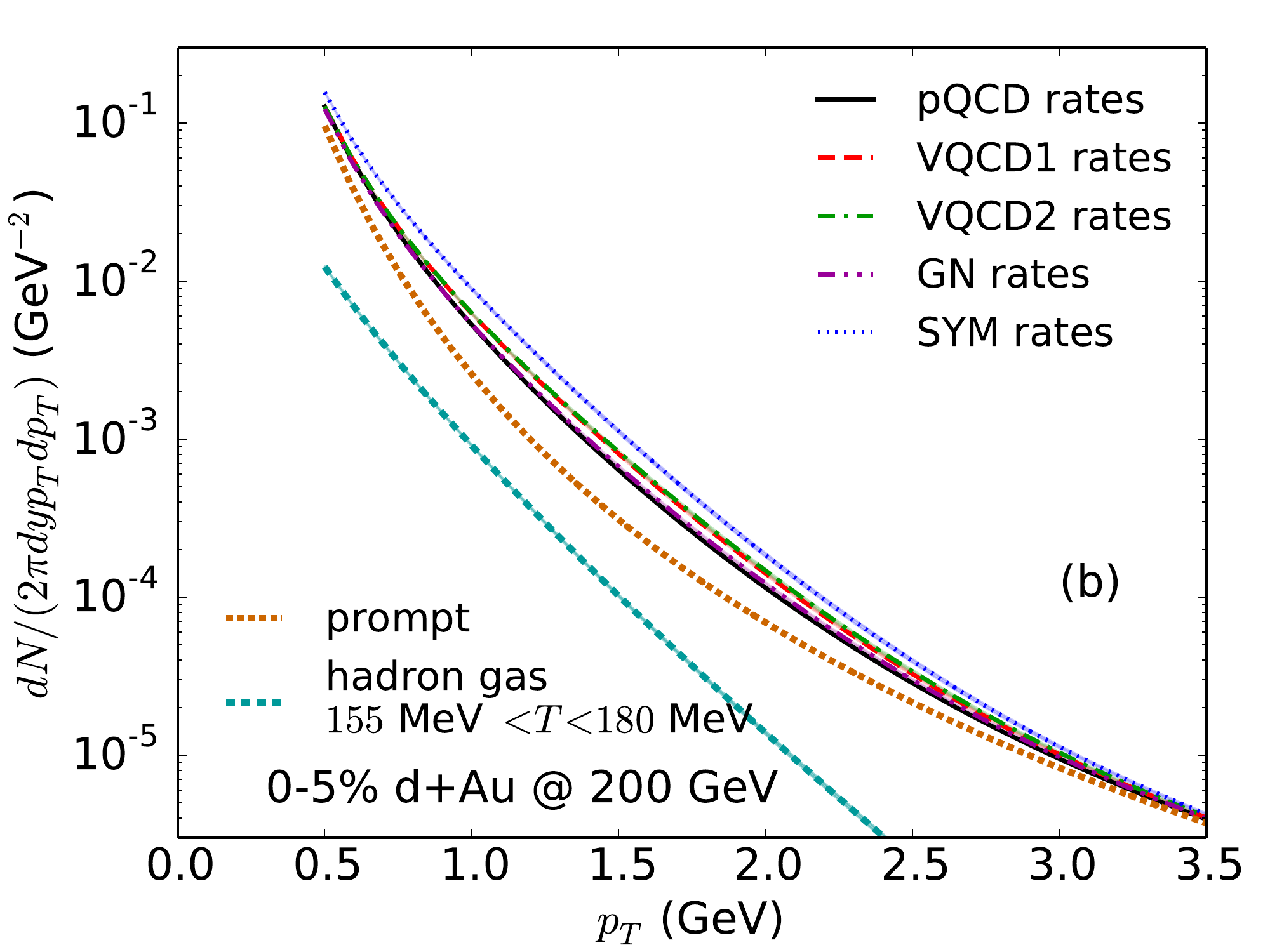} \\
		 \includegraphics[width=0.48\linewidth]{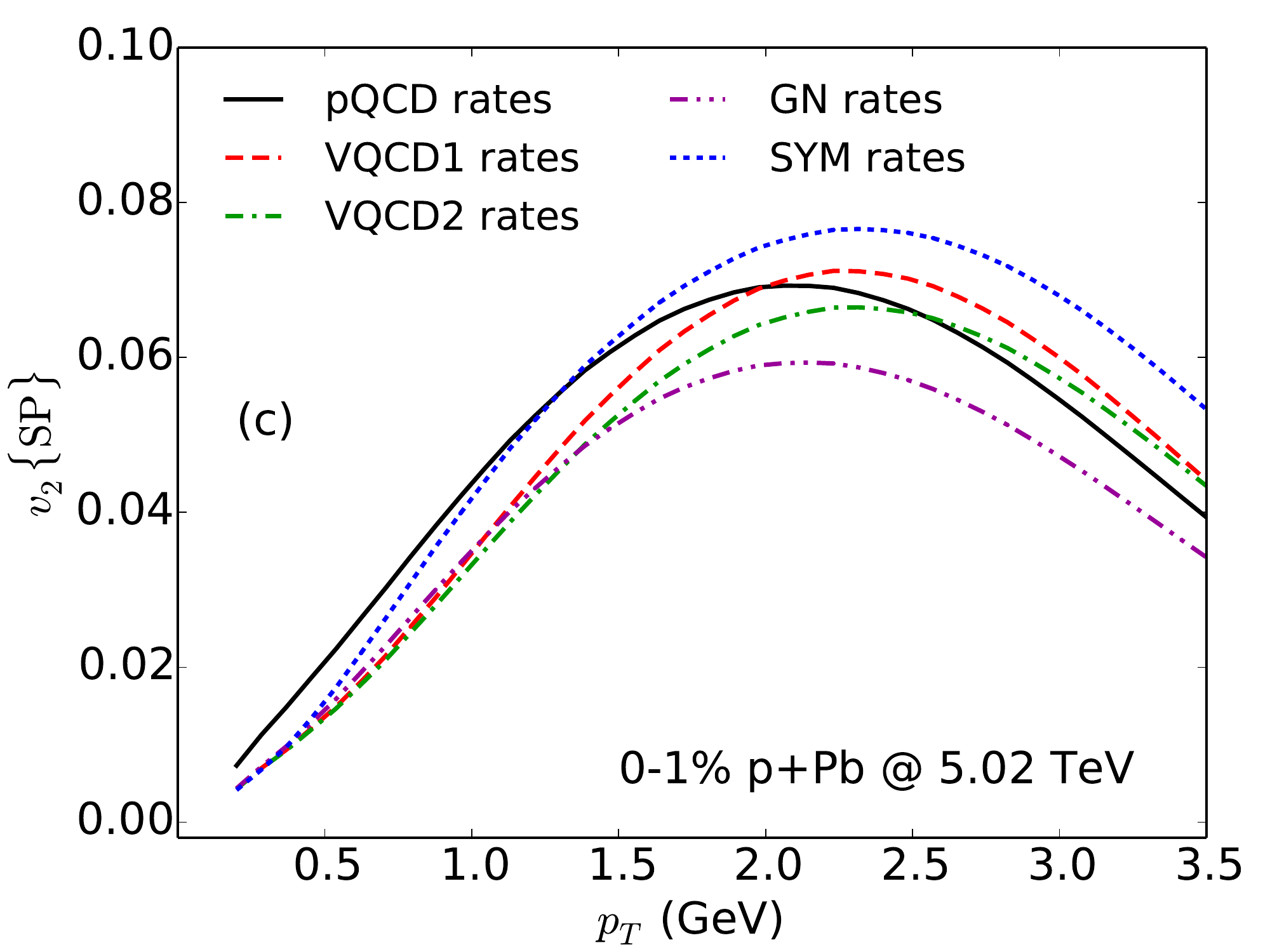} &
		 \includegraphics[width=0.48\linewidth]{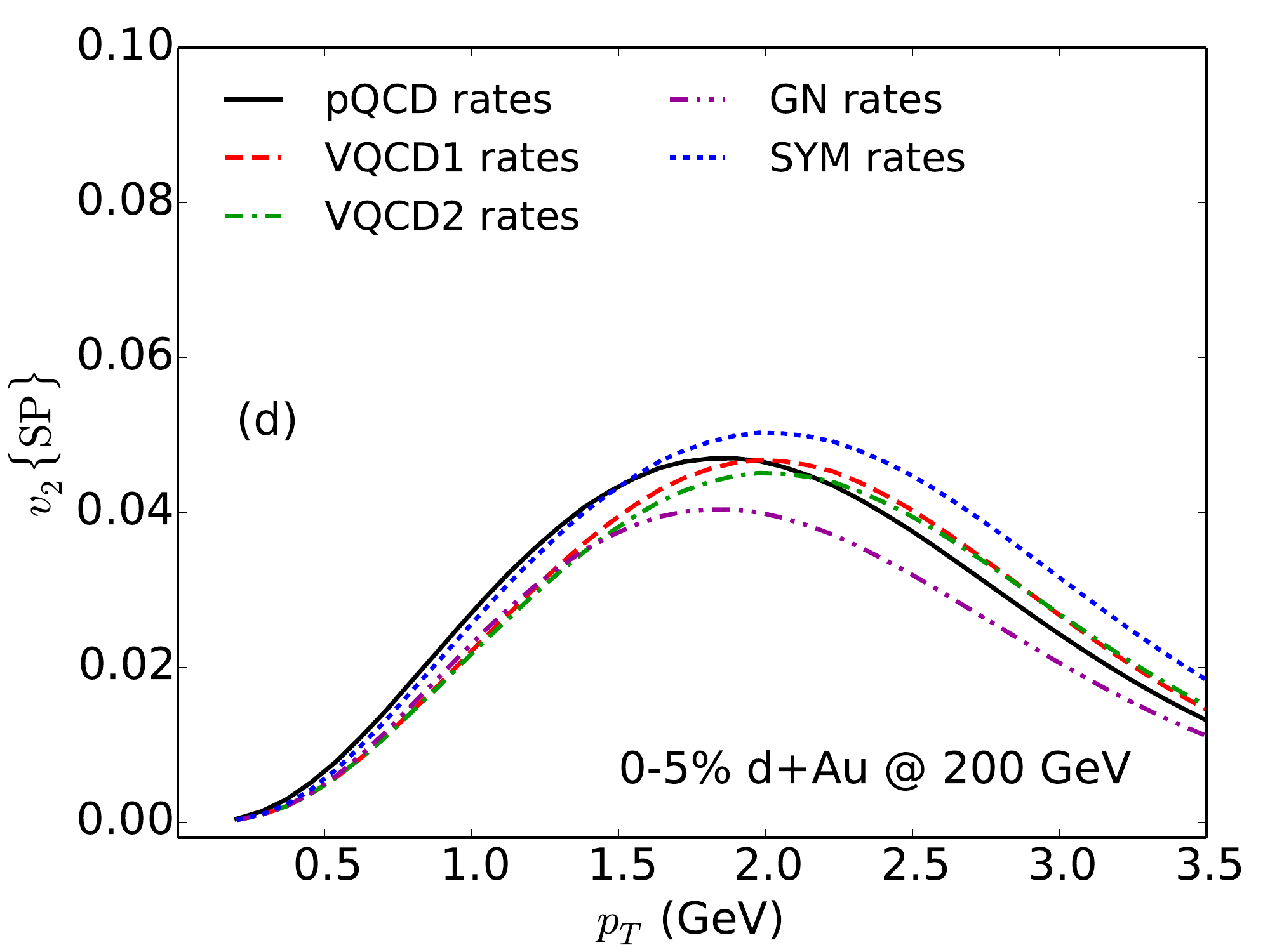} \\
		 \includegraphics[width=0.48\linewidth]{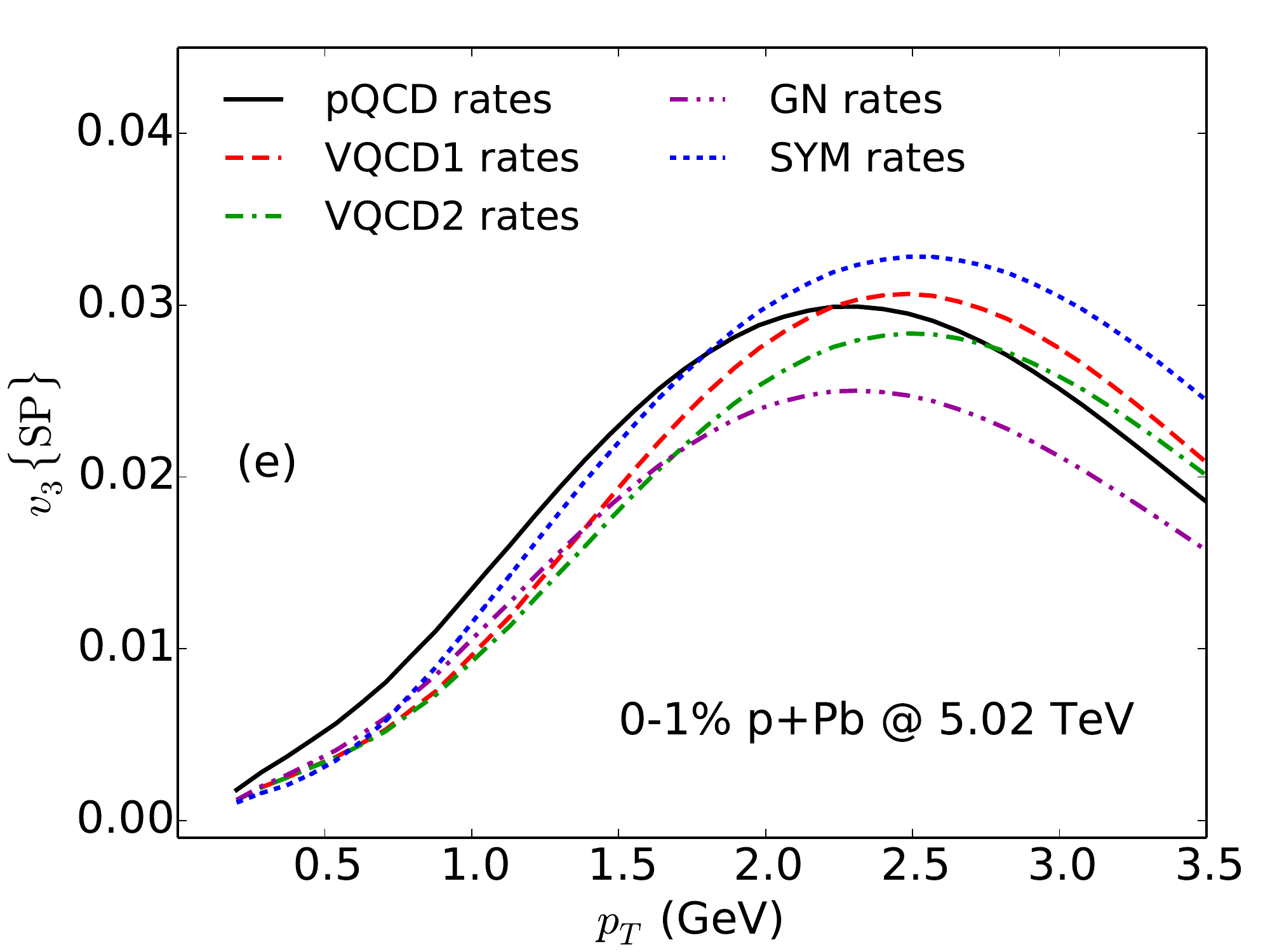} &
		 \includegraphics[width=0.48\linewidth]{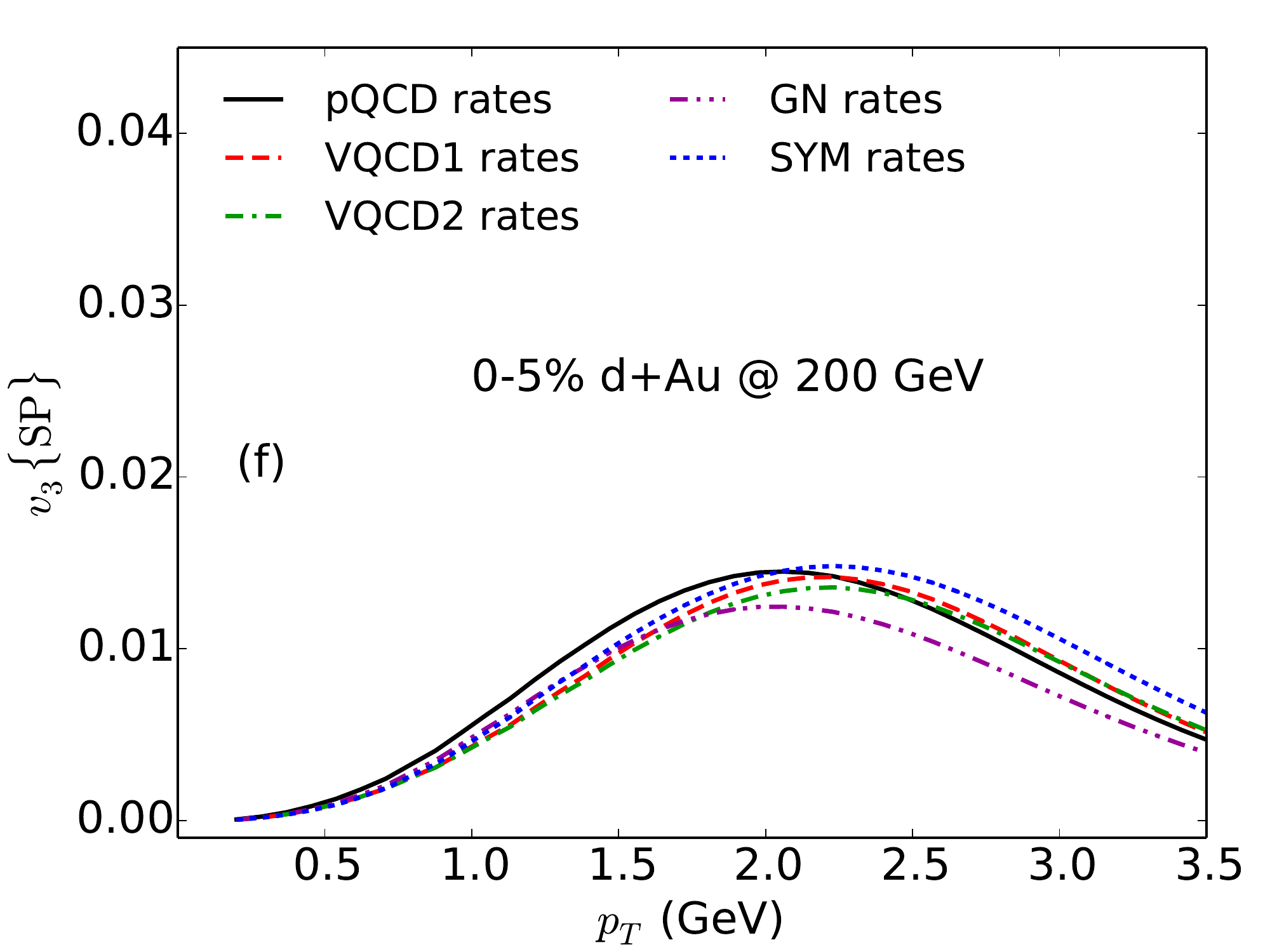}
	\end{tabular}
	\caption{(Color online) Direct photon spectra from 0-1\% p+Pb collisions at 5.02 TeV $(a)$ and 0-5\% d+Au collisions at 200 GeV $(b)$ using different sets of emission rate in the QGP phase. Direct photon anisotropic flow coefficients, $v_{2,3}(p_T)$, in 0-1\% p+Pb collisions at 5.02 TeV $(c,e)$ and 0-5\% d+Au collisions at 200 GeV $(d,f)$ using different sets of emission rate in the QGP phase.}
	\label{fig6.4}
\end{figure*}
%=======================================
%

In Figs.~\ref{fig6.4}, direct photon spectra and their anisotropic coefficients are shown for top 0-1\% p+Pb collisions at 5.02 TeV and 0-5\% d+Au collisions at 200 GeV. In high-multiplicity events of small collision systems, thermal radiation can reach up to  a factor of 2 of the prompt contribution. Similar to nucleus-nucleus collisions, the direct photon spectra using the emission rates that are derived from strongly coupled theory are larger than the QCD rates. 

The difference is smaller in the d+Au collisions compared to p+Pb collisions at the higher collision energy. Although the hierarchy of direct photon anisotropic flow coefficients remains the same as those in nucleus-nucleus collisions, the splittings among the results using different emission rates are larger in 0-1\% p+Pb collisions at 5.02 TeV. The direct photon anisotropic flow coefficients in small collision systems show a stronge sensitivity to the QGP photon emission rates.

\subsection{Discussion}

In general, direct-photon production with holographic models is enhanced compared to the one with pQCD especially in intermediate $p_T$, which may stem from the blue shift of thermal-photon-emission rates in strongly coupled scenarios as shown in Section \ref{sec_photon_emission}, which is also in accordance with Fig.\ref{fig6.1.2}. On the other hand, we find that the anisotropic flow of direct photons with holographic models is generally suppressed in low $p_T$. On the contrary, in high $p_T$, the flow from V-QCD and the SYM plasma surpasses the one from pQCD. To analyze the hierarchy of the flow of direct photons by changing sources in the QGP phase, we further illustrate the flow of ``thermal photons" generated in the QGP phase, where the photons from hadronic phase are excluded, from distinct sources.

\begin{figure}[ht!]
	\begin{minipage}{7.5cm}
		\begin{center}
			 \includegraphics[width=7.5cm,height=6cm,clip]{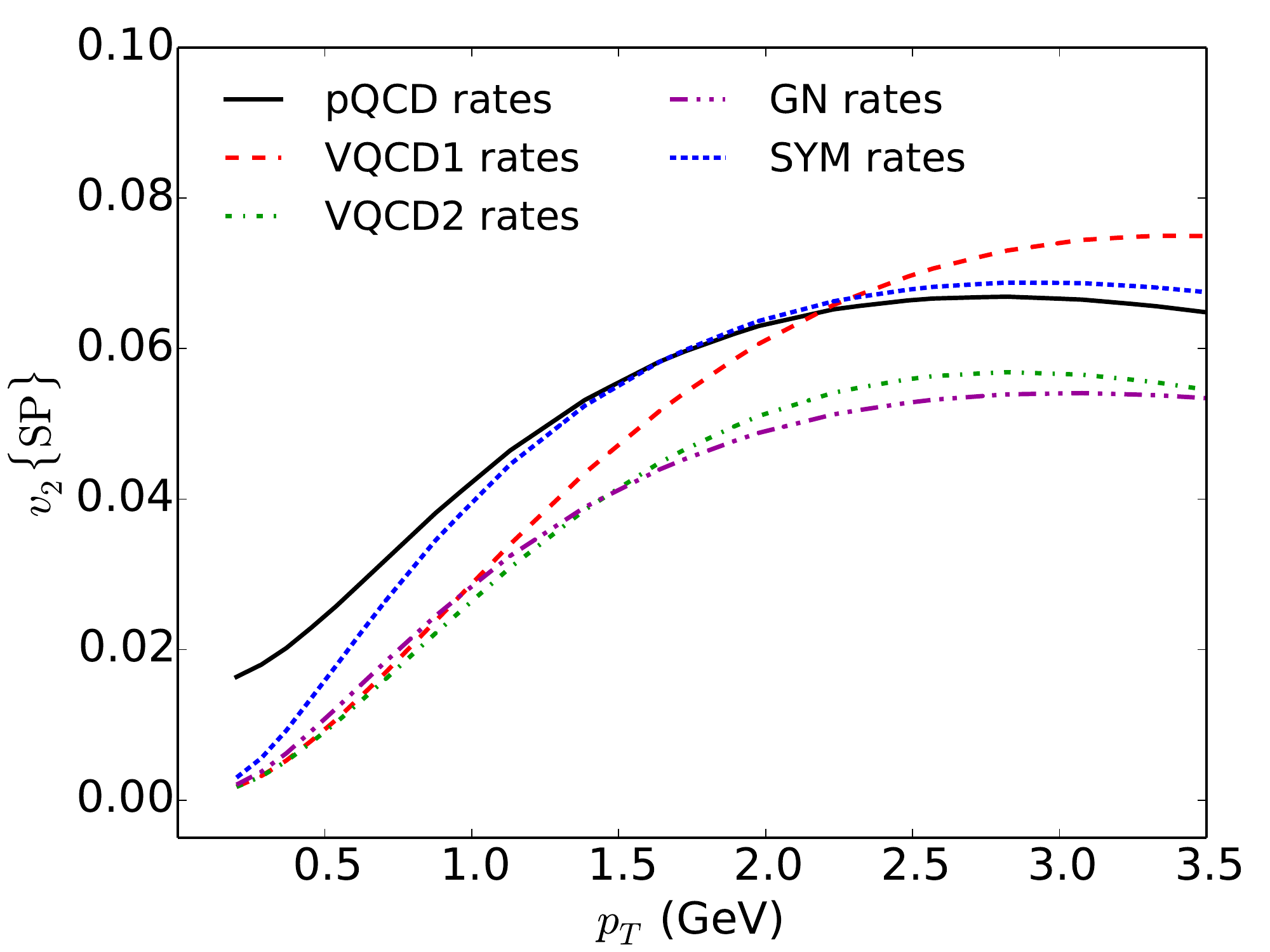}
			\caption{(Color online) Thermal QGP photon $v_2(p_T)$ using different emission rates in the QGP phase in 0-40\% Pb+Pb collisions at 2.76 $A$\,TeV.}
			\label{fig6.7}
		\end{center}
	\end{minipage}
	\hspace {1cm}
	\begin{minipage}{7.5cm}
		\begin{center}
			{\includegraphics[width=7.5cm,height=6cm,clip]{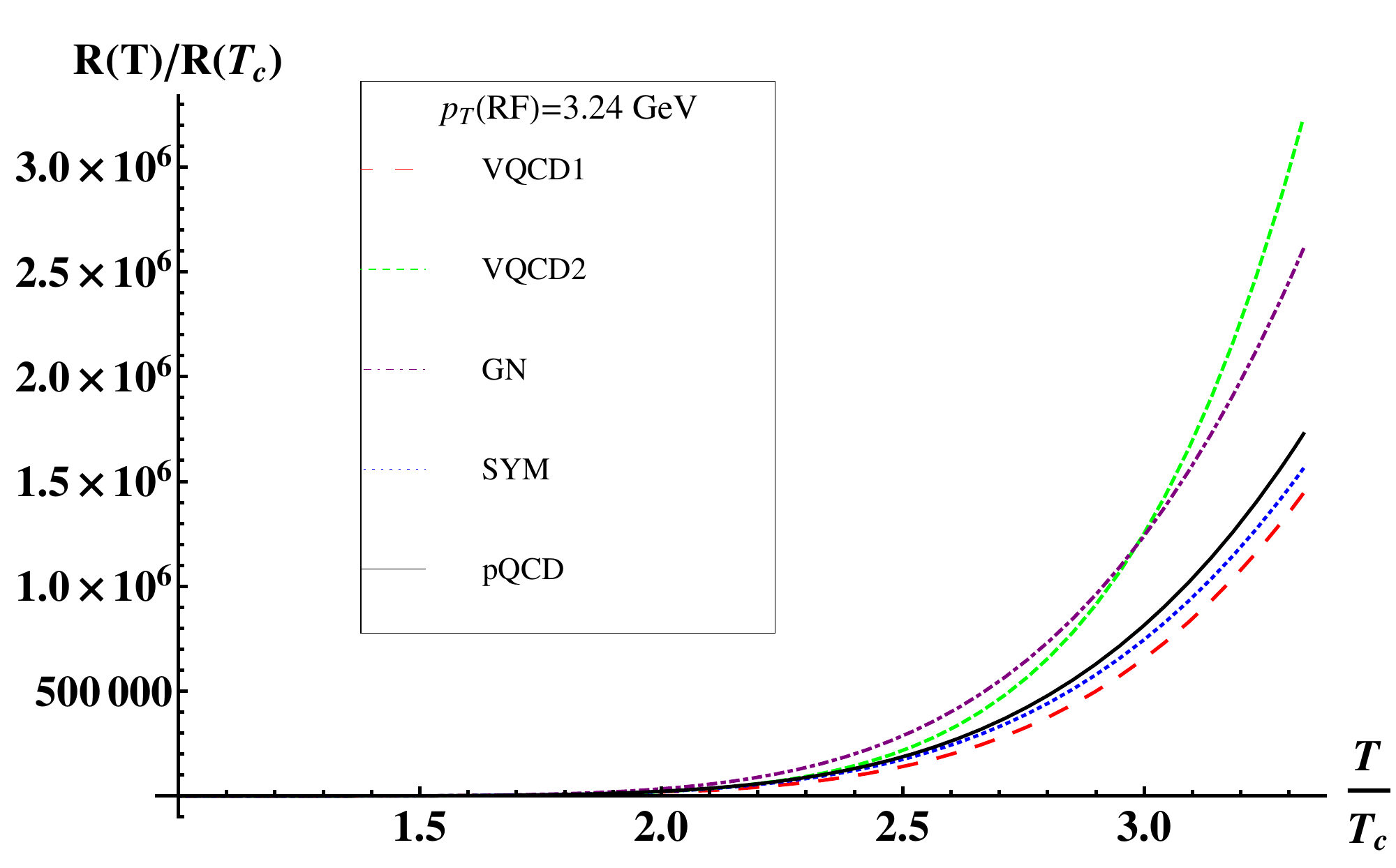}}
			\caption{Ratios of emission rates versus temperature at high $p_T=3.24$ GeV in the rest frame(RF). }\label{R_ratio_high_pt}
		\end{center}
	\end{minipage}
	\hspace {1cm}
\end{figure}

We firstly focus on nucleus-nucleus collisions especially for the LHC collisional energy. As shown in Fig.\ref{fig6.7}, except for the one from pQCD, the hierarchy of flow for "thermal photons" is similar to the one for direct photons in high $p_T$ as shown in Fig.\ref{fig6.3}. Since the photon emission at $p_T \in [2, 4]$ GeV from the QGP phase dominates the one from hadronic phase as illustrated in Fig.\ref{fig6.1.2}, the momentum anisotropy of the thermal photons from the QGP phase is the only cause of flow. The high $p_T$ regime thus allows us to unambiguously compare the effects on flow from different sources without the contamination from hadronic contributions.

It turns out that pQCD, SYM, and VQCD1 lead to larger flow of thermal photons from the QGP phase for both $v_2$ and $v_3$ in this region as shown in Fig.\ref{fig6.7}. Since the flow is related to the weight of photon emission at different temperature (equivalently the time elapse) during the expansion of the medium, we may illustrate the ratios of the emission rates at $T$ to the ones at $T_c$ in high $p_T$ to have an intuitive understanding of the hierarchy of flow. As shown in Fig.\ref{R_ratio_high_pt}, the high-$p_T$ photons are dominantly generated at high temperature\footnote{The fixed $p_T$ in Fig.\ref{R_ratio_high_pt} is in the rest frame of the fluid cell, which in fact differs from the $p_T$ in the lab frame, while the analysis may still capture "qualitative" features of the flow at high $p_T$ in the lab frame.}. The smaller ratios of emission rates in high temperature should result in larger flow due to larger weights of photons emitted at lower temperature. The hierarchy of emission-rate ratios at high temperature in Fig.\ref{R_ratio_high_pt} thus approximately manifests the hierarchy of thermal-photon flow at high $p_T$ in Fig.\ref{fig6.7}.

It is worthwhile to note here that the hierarchy of the spectral functions in high energies could be distinct from the hierarchy in low energies. Nevertheless, for the direct-photon flow, one finds that the one for pQCD rate is suppressed at high $p_T$ as shown in Fig.\ref{fig6.3}, which contradicts the thermal-photon flow from the QGP phase. The suppression of flow for pQCD stems from the dilution by prompt photons, while the dilution for holographic models is less severe due to larger rates of holographic models at high $p_T$. In conclusion, the direct-photon flow at high $p_T$ is governed by the competition between the dilution by prompt photons and the ratios of emission rates of thermal photons in the QGP phase at high temperature.

On the other hand, at low $p_T$ or intermediate $p_T$, the thermal photons from hadronic phase become more significant, which play a central role for the direct-photon flow. Due to the interplay between thermal photons created in QGP and hadronic phases, it is more difficult to analyze the hierarchy of flow from distinct models. Nonetheless, the enhancement of the direct-photon spectra for holographic models at low $p_T$ is minor. The discrepancies between theoretical predictions and experimental observations in spectra and flow at low $p_T$ are more associated with underestimation of thermal photons from hadronic radiation in our study.

In the simulations for small collision systems, due to the suppression of thermal photons from the confined phase as shown in (a) and (b) of Fig.\ref{fig6.4}, the flow is primarily engendered by photons from the QGP phase. In addition, one finds that the thermal-photon spectra are more dominant than the prompt-photons spectrum in the $0-1\%$ p+Pb collisions, whereas the thermal-photon spectra become more comparable to the prompt-photons spectrum in the $0-5\%$ d+Au collisions. The direct-photon flow in the latter case is accordingly suppressed in comparison with the former case  as illustrated in (c)-(f) of Fig.\ref{fig6.4}. Moreover, since the spectra from holographic models are more prominent than the one from pQCD in the former case, we find that the flow for holographic models dominates the flow for pQCD at high $p_T$. 

However, in the latter case, the flow from different models are more comparable due to their approximate spectra at high $p_T$. In intermediate $p_T$, because of the considerable suppression of prompt photons, the flow reaches maximum in this region. At low $p_T$, photons from hadron gas may start to affect the flow though their influence is highly suppressed compared to the case in AA collisions. On the other hand, since the spectra for different holographic models and pQCD converge, particularly in the $0-5\%$ d+Au collisions, the splitting of flow is also reduced. Future comparisons with experimental observations in small collision systems could further verify the importance of the influence from thermal photons in the QGP phase on anisotropic flow.

\section{Concluding Remarks}\label{sec_con}

In this work, we have evaluated the direct-photon production in heavy ion collisions from holography convoluted with the medium evolution and the inclusion of other sources.
 The agreement with experiments in spectra is improved compared with the previous study by using the pQCD rate. On the other hand, the deviation in flow is increased at low $p_T$ but decreased at high $p_T$. 

In small collision systems, where the experimental data of direct photons have not been available, holographic models lead to enhancements in both spectra and flow. Our findings may highlight the strong influence of thermal photons from the QGP phase on the direct-photon flow at high $p_T$, where hadronic contributions are highly suppressed. The enhancement of flow in this region stems from the amplification of the weight of late-time emission and the amplitude of thermal-photon emission in the QGP phase. 

In our study, the blue-shift due to the increase of couplings could in general lead to the latter effect, whereas the former one is in fact model dependent. Furthermore, as anticipated, our study also suggests that the hadronic contributions are responsible for the flow
at low $p_T$. In contrast to the nucleus-nucleus collisions, the dominance of QGP photons could be more pronounced in small collision systems for larger $p_T$ window. Therefore, future measurements of direct-photon spectra and flow in small systems will be crucial to understand the electromagnetic property of the sQGP. Moreover, direct-photon production at high $p_T$ is sensitive to initial conditions with the matching to hydrodynamics and other sources such as photons from non-equilibrium states \cite{Benic:2016uku,Greif:2016jeb,Oliva:2017pri,Berges:2017eom} or the radiation of jets \cite{Turbide:2007mi,Greif:2016jeb}. Furthermore, the thermal-photon emission rate itself could be modified by viscosities \cite{Paquet:2015lta}. A more comprehensive study incorporating all other sources should be pursued in the future.

\begin{acknowledgments}

I.I. is grateful to I. Zahed for useful discussions. E.K. would like to thank B. Erazmus and Y. Foka for useful conversations. D. Y. would like to thank M. Huang and D. Li for fruitful discussions in the early stage of this work.

The work of I.I. is part of the D-ITP consortium, a program of the Netherlands Organisation for Scientific Research (NWO) that is funded by the Dutch Ministry of Education, Culture and Science. The work of E. K. was partially supported by  European Union's Seventh Framework Programme under grant agreements (FP7-REGPOT-2012-2013-1) no 316165 and the Advanced ERC grant SM-grav, No 669288. D.Y. was supported by the RIKEN Foreign Postdoctoral Researcher program. C. S. was supported in part by the U.S. Department of Energy, Office of Science under contract No. DE-SC0012704 and the Natural Sciences and Engineering Research Council of Canada. C.S. gratefully acknowledges a Goldhaber Distinguished Fellowship from Brookhaven Science Associates.
Computations were made in part on the supercomputer Guillimin from McGill University, managed by Calcul Qu\'ebec and Compute Canada. The operation of this supercomputer is funded by the Canada Foundation for Innovation (CFI), NanoQu\'ebec, RMGA and the Fonds de recherche du Qu\'ebec - Nature et technologies (FRQ-NT). 
\end{acknowledgments}

\begin{figure}[ht!]
	\begin{minipage}{7.5cm}
		\begin{center}
			\includegraphics[width=8.5cm,height=6cm,clip]{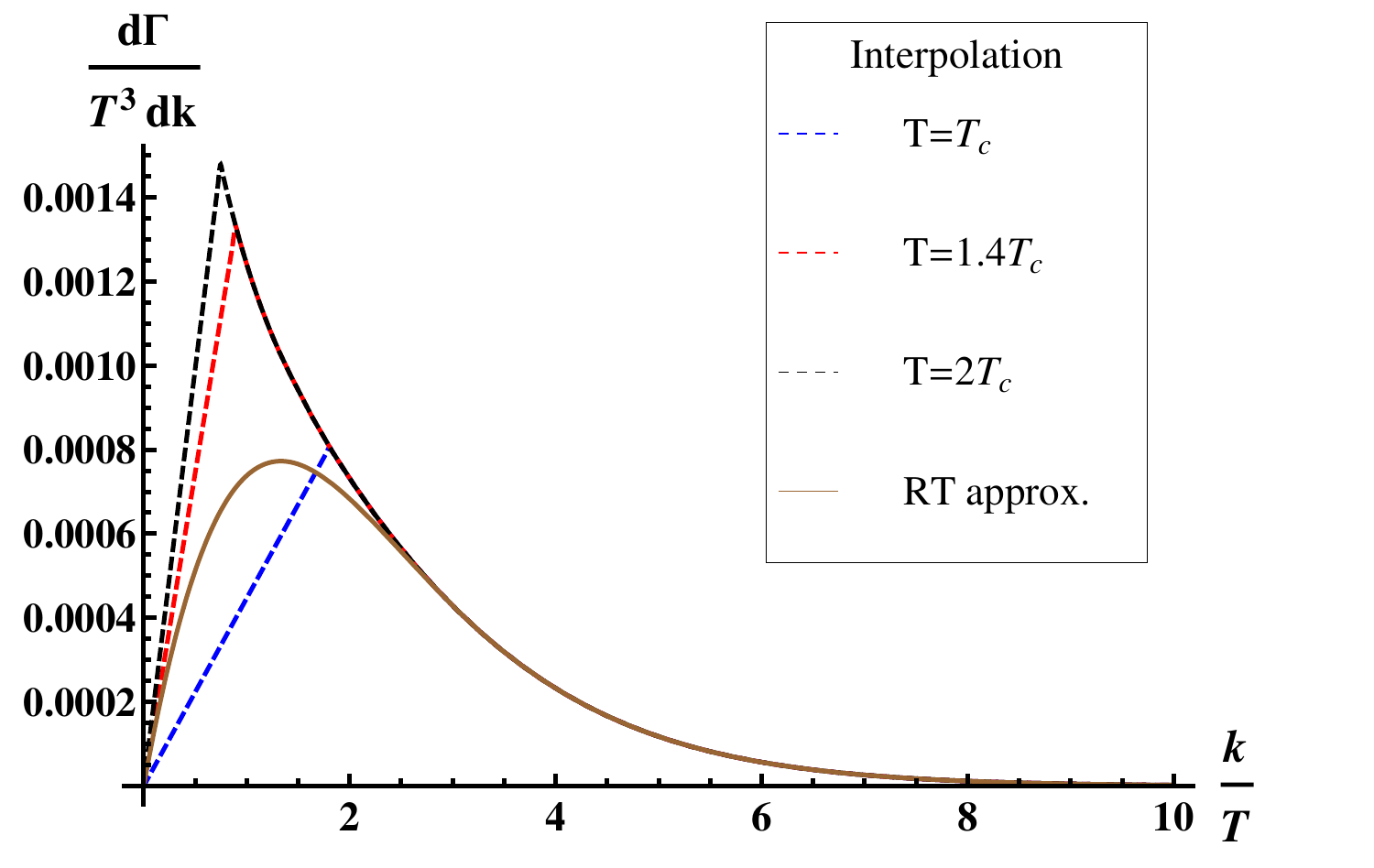}
			\caption{Thermal-photon emission rates with different interpolations.}
			\label{interpolation_rate_2}
		\end{center}
	\end{minipage}
	\hspace {1cm}
	\begin{minipage}{7.5cm}
		\begin{center}
			{\includegraphics[width=8.5cm,height=6cm,clip]{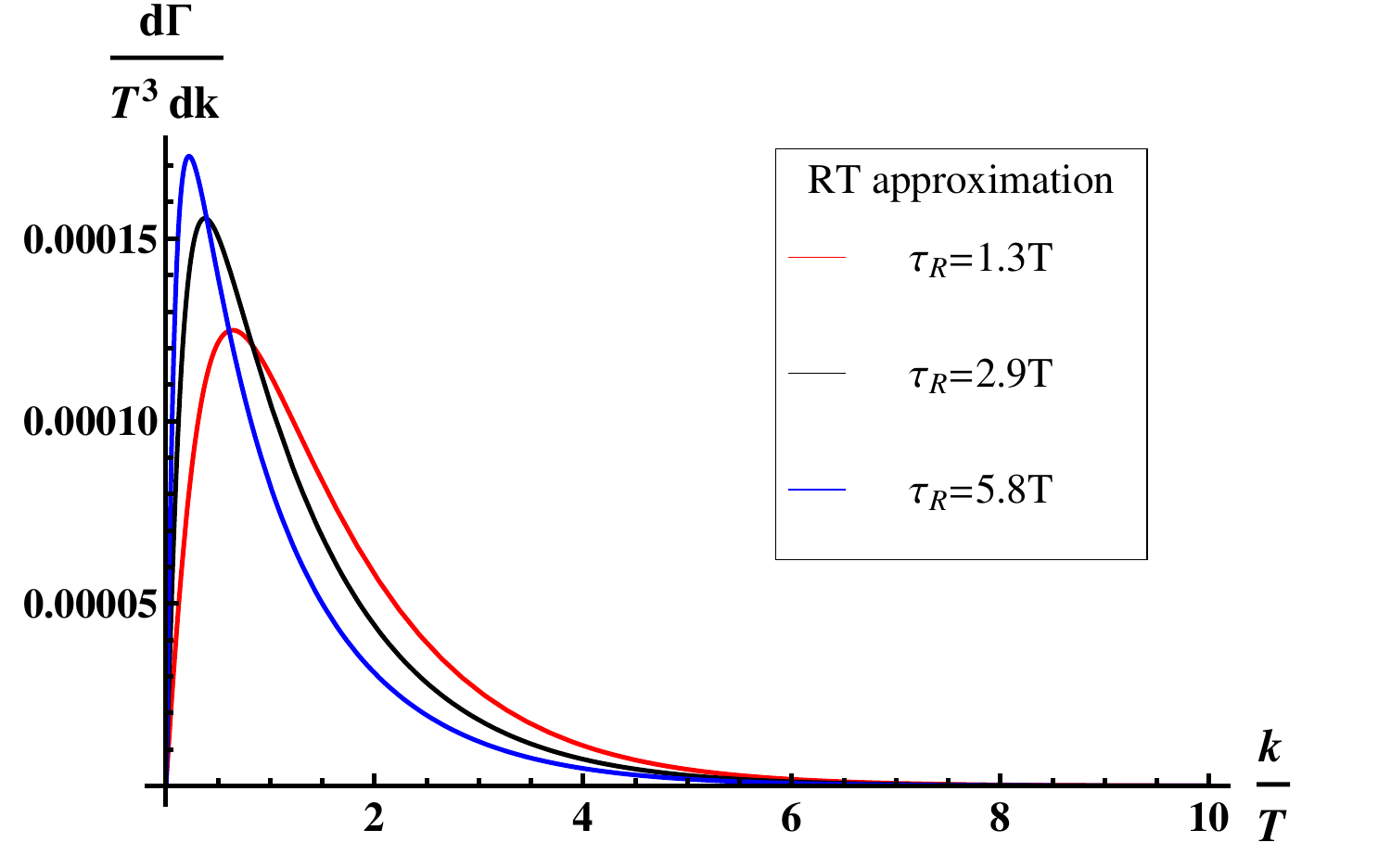}}
			\caption{Photo-emission rates from the RT approximation with different couplings (relaxation times).}\label{RT_comparison}
		\end{center}
	\end{minipage}
	\hspace {1cm}
\end{figure}   

\begin{figure}[ht!]
	\centering
	\includegraphics[width=0.48\linewidth]{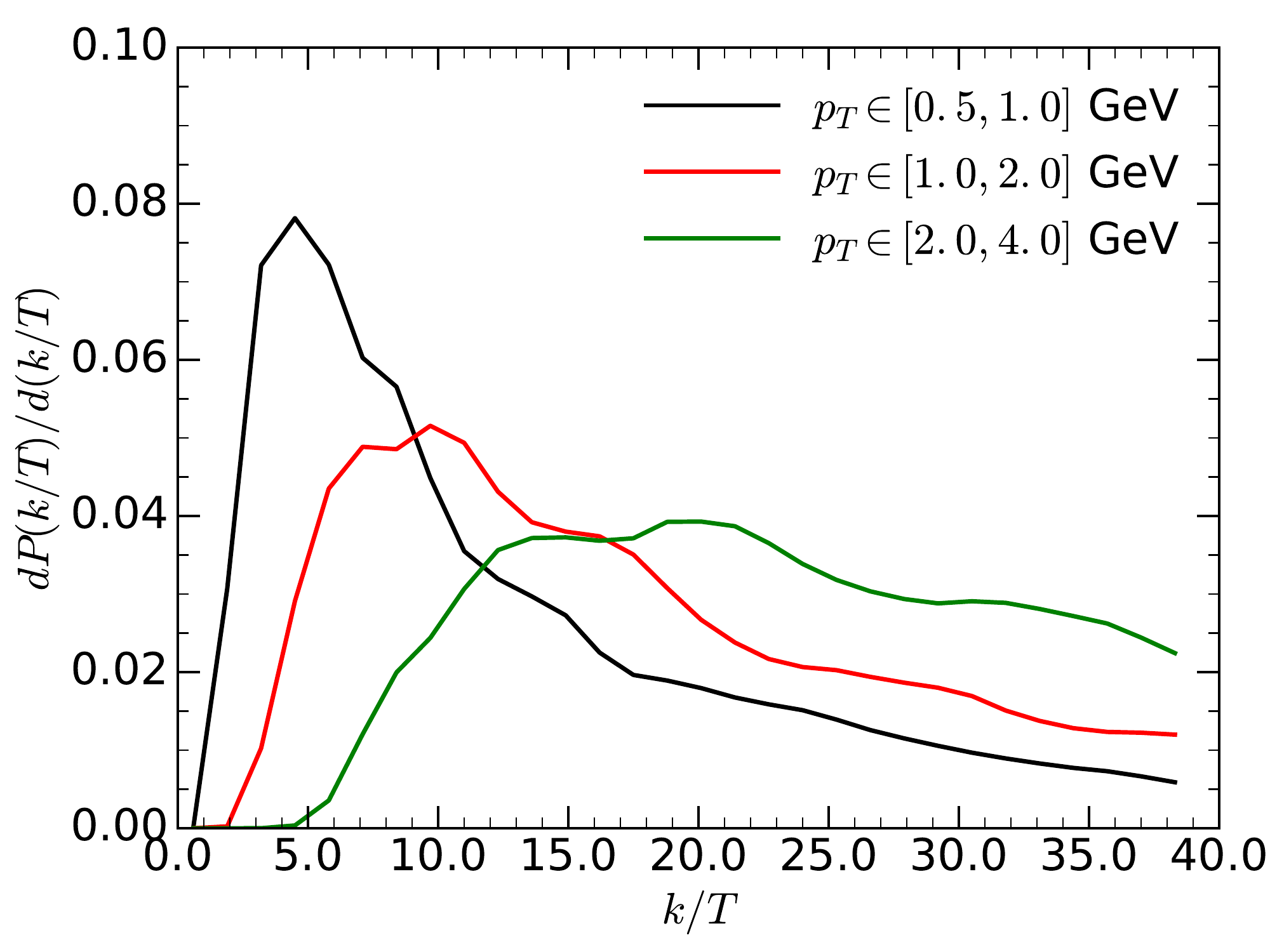}
	\caption{The probability distribution of the values of $k/T$ at which the QGP rates were evaluated in a typical 0-40\% Pb+Pb collisions at 2.76 TeV. The three curves represent the probability distributions in three different photon transverse momentum ($p_T$) ranges in the lab frame. }
	\label{fig.koverT_stat}
\end{figure}

\begin{figure*}[ht!]
	\centering
	\begin{tabular}{cc}
		\includegraphics[width=0.48\linewidth,height=0.58\linewidth]{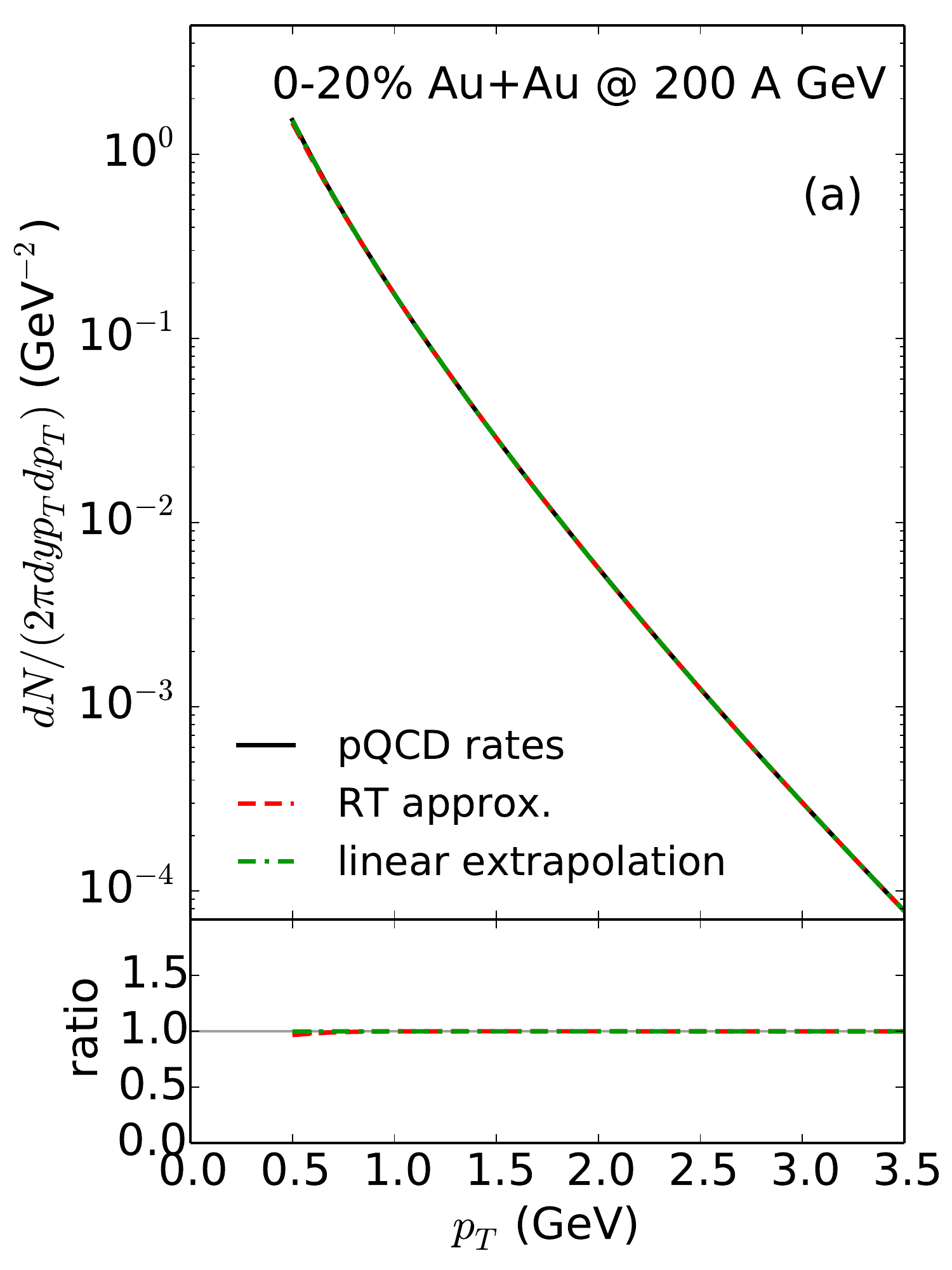} &
		\includegraphics[width=0.48\linewidth,height=0.58\linewidth]{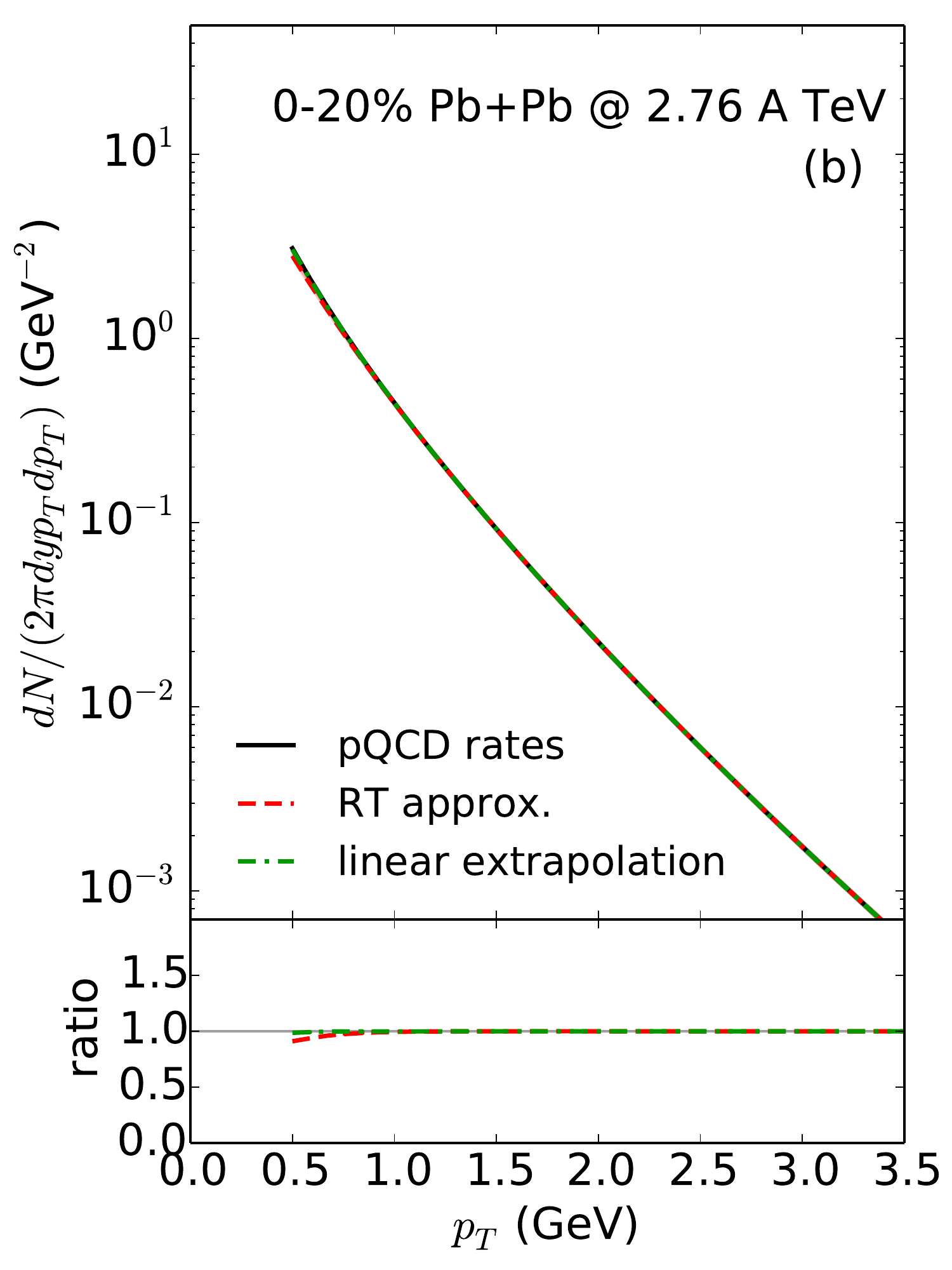} \\
		\includegraphics[width=0.48\linewidth,height=0.58\linewidth]{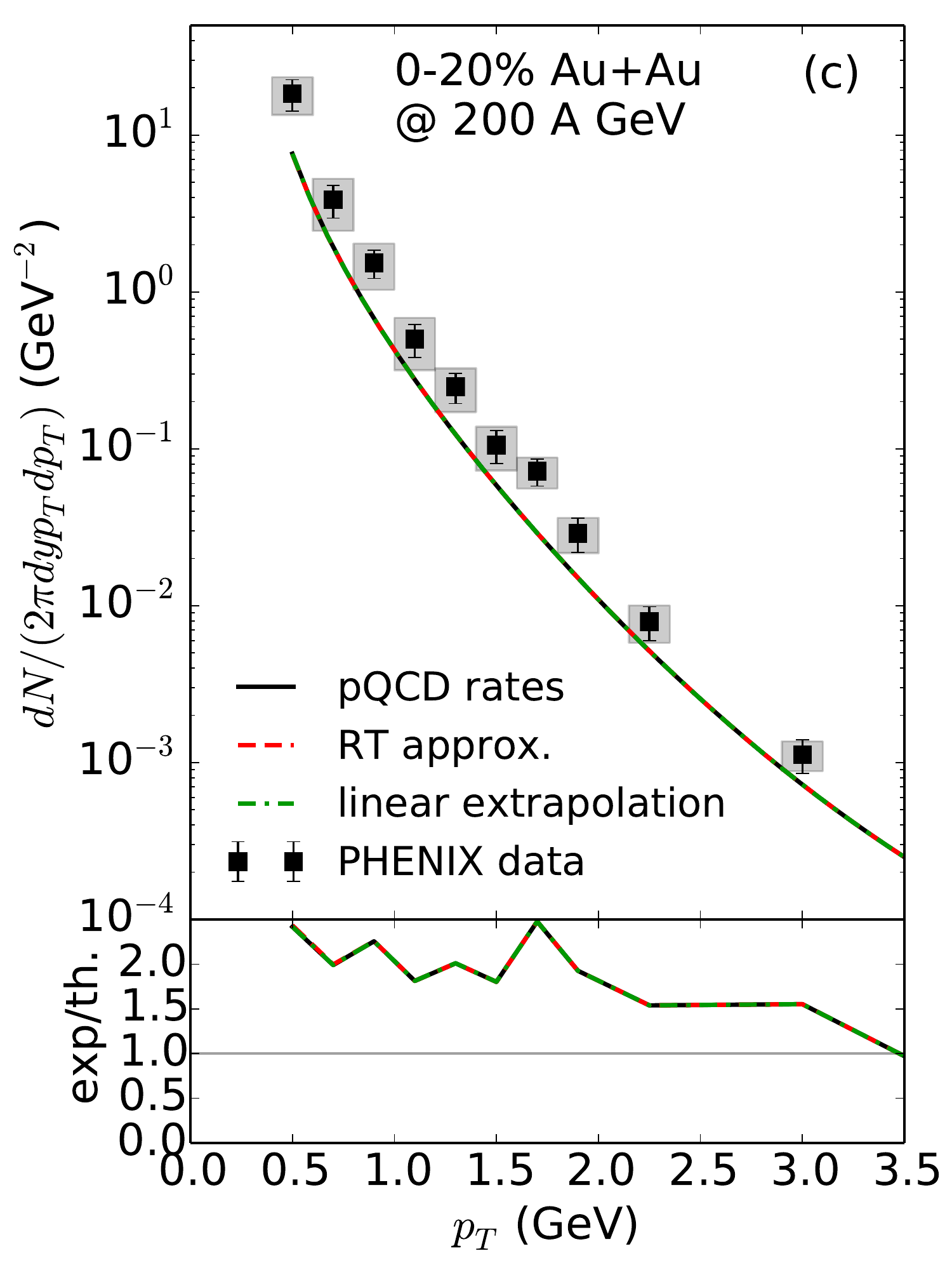} &
		\includegraphics[width=0.48\linewidth,height=0.58\linewidth]{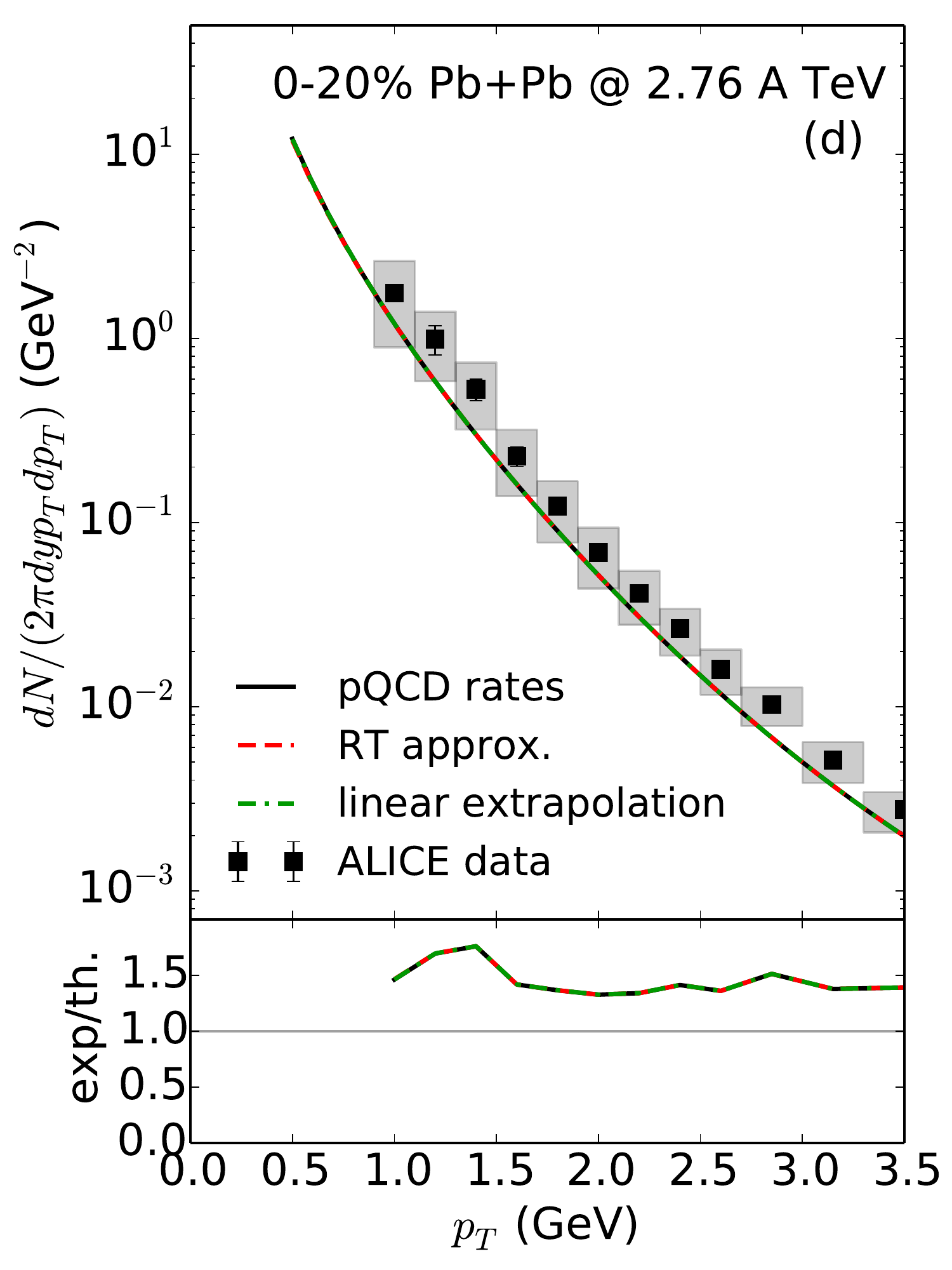}
	\end{tabular}
	\caption{QGP-photon (upper plots) and Direct-photon (lower plots) spectra led by pQCD rates with different interpolations in low energy. Black curves here denote the original results without interpolation.}
	\label{fig.int_spectra}
\end{figure*}

\begin{figure*}[ht!]
	\centering
	\begin{tabular}{cc}
		\includegraphics[width=0.48\linewidth]{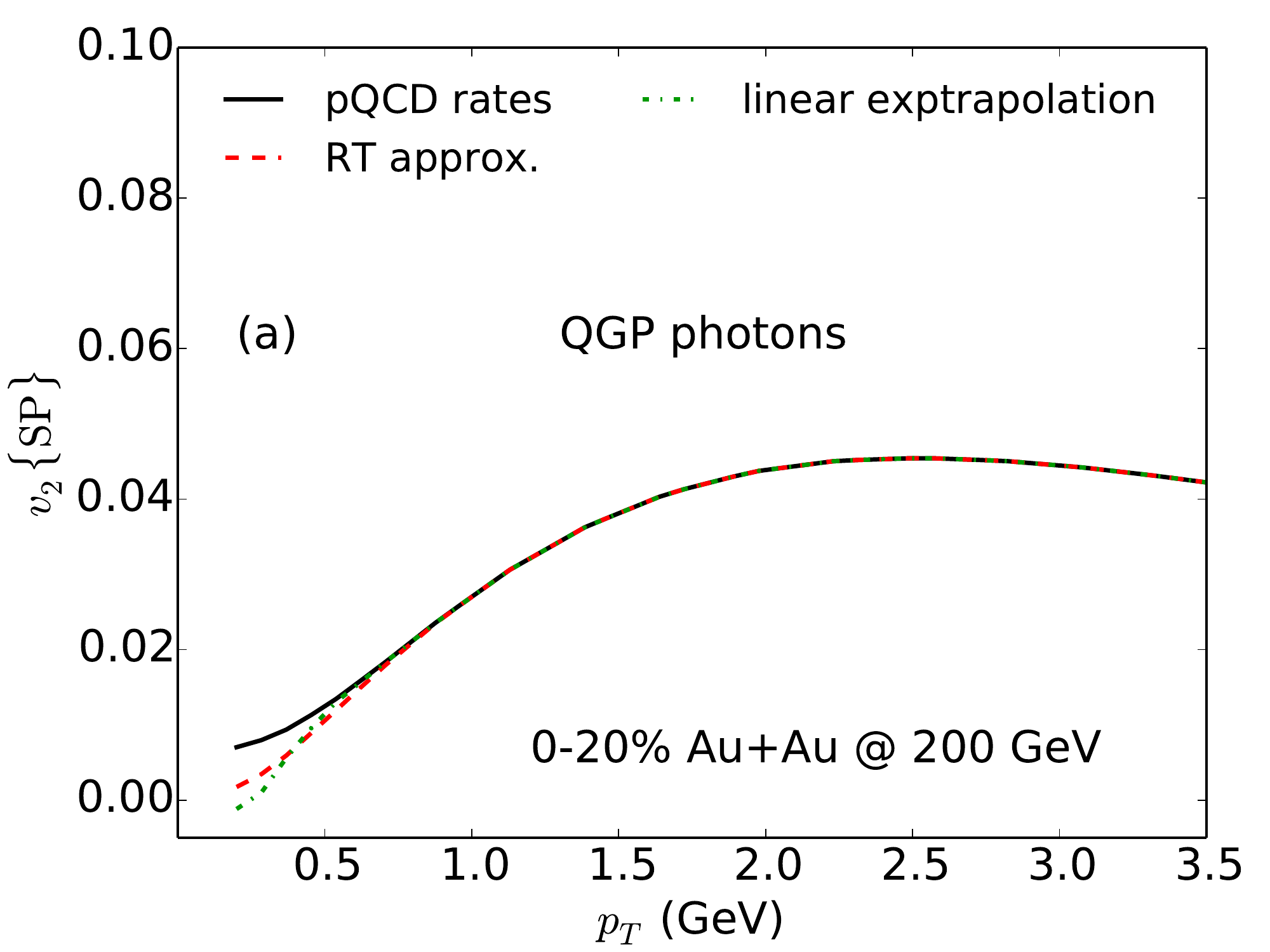} &
		\includegraphics[width=0.48\linewidth]{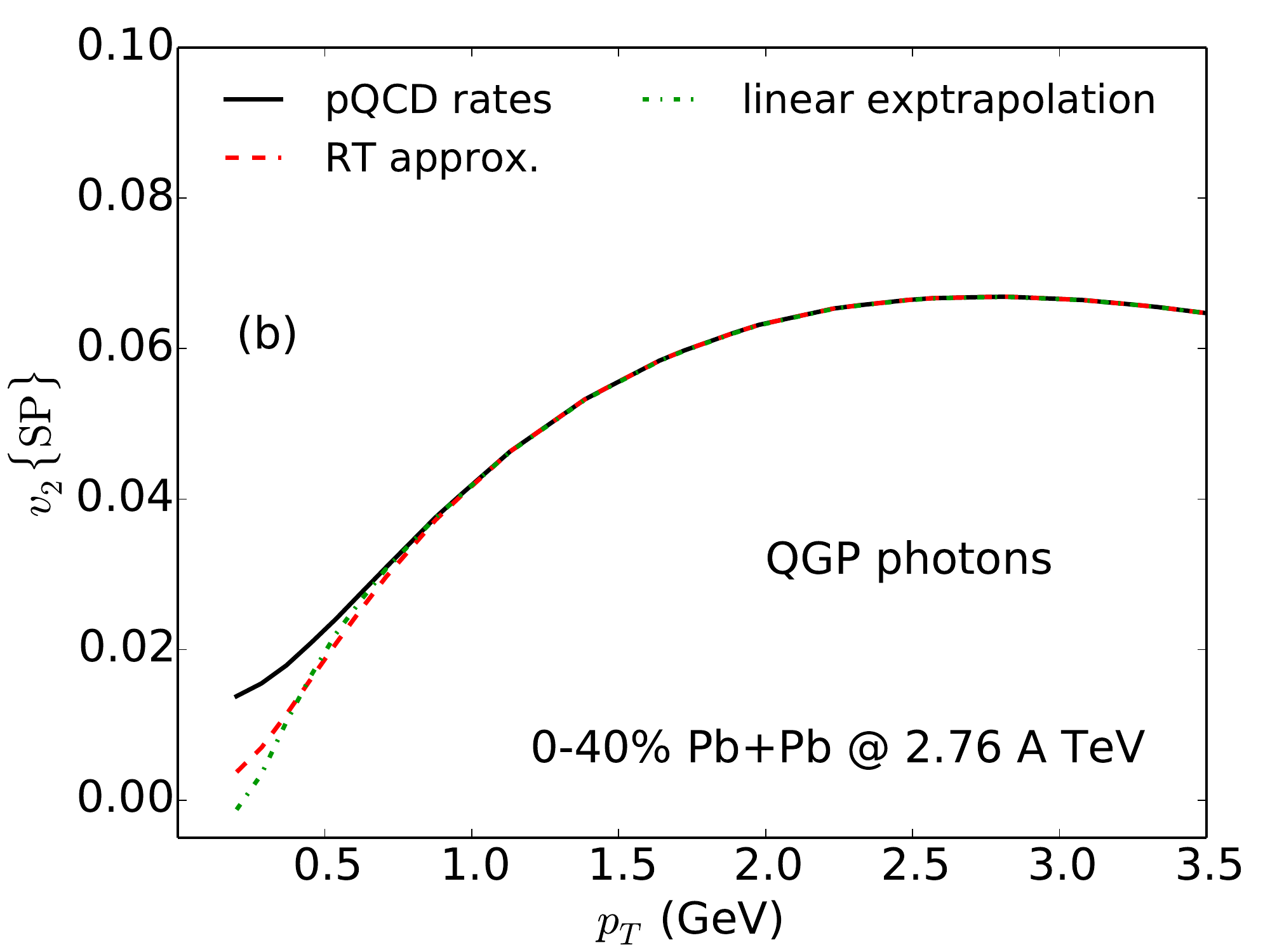} \\
		\includegraphics[width=0.48\linewidth]{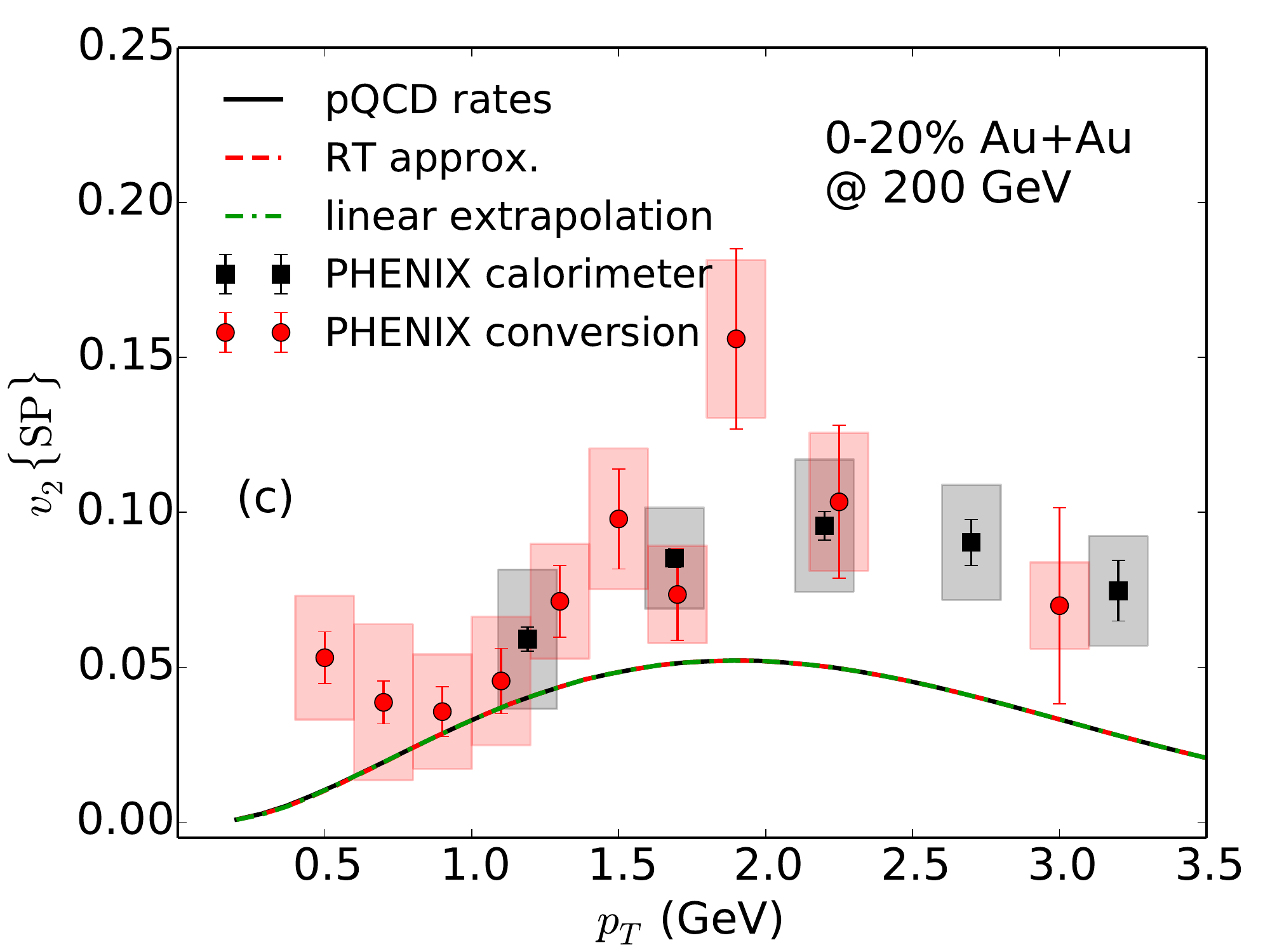} &
		\includegraphics[width=0.48\linewidth]{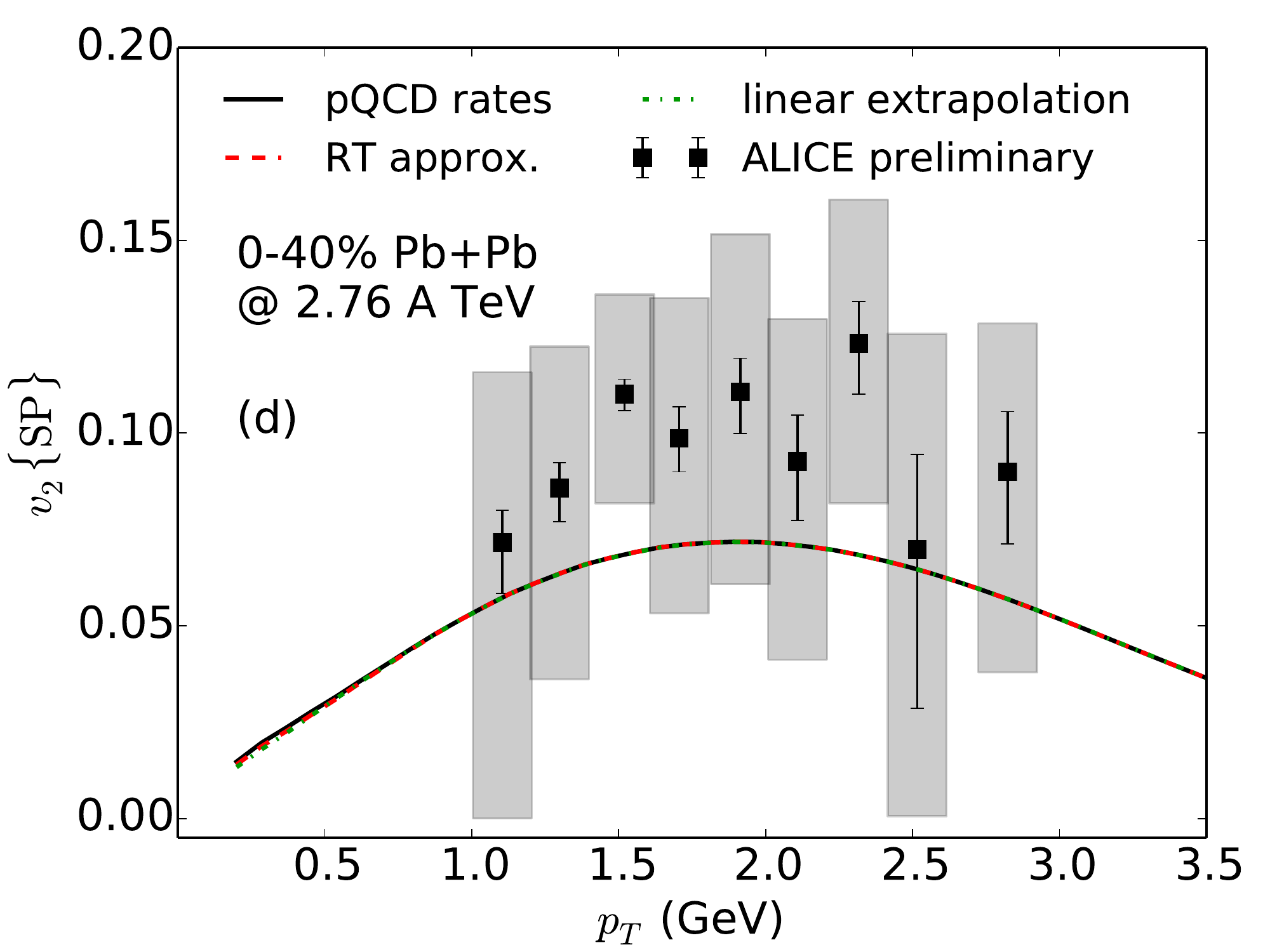}
	\end{tabular}
	\caption{QGP-photon (upper plots) and Direct-photon (lower plots) flow led by pQCD rates with different interpolations in low energy.  Black curves here denote the original results without interpolation.}
	\label{fig.int_flow}
\end{figure*}

\appendix
\section{Interpolation for Low-Energy Emission Rates}
In this Appendix, we apply both relaxation-time (RT) approximation and linear extrapolation to estimate the pQCD emission rates in small momenta. 
We follow the procedures in \cite{Romatschke:2015gic} to evaluate photon emission rates via the RT approximation.
As shown in \cite{Romatschke:2015gic}, under the perturbation of a near-equilibrium distribution function $f(x,q)$, the Boltzmann equation of quasi-particles yields
\begin{eqnarray}
v^{\mu}\partial_{\mu}\delta f-\frac{\bf v\cdot E}{T}f_0(1-f_0)=-\frac{1}{\tau_R}(\delta f-\delta f_0)
\end{eqnarray}
where $v^{\mu}=(1,{\bf q}/|{\bf q}|)$ and $f_0(q)=1/(e^{(|{\bf q}|-\mu)/T}+1)$ represents the equilibrium distribution function of fermions with $\mu$ being a chemical potential and $\tau_R$ corresponds to the relaxation time. Here ${\bf E}=\nabla A_0-\partial_t{\bf A}$ is the electric field. The small perturbation of the equilibrium distribution function comes from
\begin{eqnarray}
\delta f_0=\frac{\delta\mu}{T}f_0(1-f_0),
\end{eqnarray}
where we assume $T$ is a constant. 
By making Fourier transformation of the Boltzmann equation, one finds
\begin{eqnarray}\label{delta_f}
\delta f(k,q)=\frac{f_0(q)}{T}(1-f_0(q))\frac{\delta\mu+\tau_R{\bf v\cdot E}}{1+i\tau_R k\cdot v}.
\end{eqnarray} 
Considering the fluctuation of current density $\delta J^{\mu}(t,x^3)$, which is equivalent to take $k^{\mu}=(-\omega,0,0,k)$, we are now interested in just retarded correlators of transverse currents for $k=\omega$. By using (\ref{delta_f}), we find
\begin{eqnarray}\nonumber
\delta J^1(w,k)&=&\int \frac{d^3q}{(2\pi)^3}v^1\delta f(q,k)
=\int\frac{d q}{2\pi^2T}q^2f_0(q)(1-f_0(q))\int \frac{d\Omega}{4\pi} \frac{v^1(\delta\mu+\tau_R{\bf v\cdot E})}{1+i\tau_R(-w+v^3k)}.
\end{eqnarray}
From \cite{Romatschke:2015gic}, through the linear-response theory $G^R_{1,1}(w,k)=\frac{\delta J^1/{Q_e}}{\delta A^1}$ with $Q_e$ being the charge of particles, it is found that the retarded correlators 
\begin{eqnarray}
G^R_{1,1}(\omega,k)=\frac{i\chi \omega \tau_R}{2}\left(\frac{(1-i\omega\tau_R)}{k^2\tau_R^2}+
\frac{i\left(1-2i \omega \tau_R-\tau_R^2\omega^2\right)+i k^2\tau_R^2}{2\tau_R^3k^3}\log\left[\frac{\omega-k+\frac{i}{\tau_R}}{\omega+k+\frac{i}{\tau_R}}\right]
\right),
\end{eqnarray}
where 
\begin{eqnarray}
\chi=\int\frac{d q}{2\pi^2 T}q^2f_0(q)(1-f_0(q))
\end{eqnarray}
corresponds to the static susceptibility. Note that $G^R_{1,1}(\omega,k)=G^R_{2,2}(\omega,k)$ by symmetry. Since we are interested in photo-emission rates, only the imaginary part of the light-like correlator is relevant,
	\begin{eqnarray}
	\text{Im}\left[G^R_{1,1}(\omega,\omega)\right]=\chi\left(\frac{1}{2\tau_{R} \omega}-\frac{1}{4\tau_R^2\omega^2}\tan^{-1}[2 \tau_R \omega]-\frac{\tau_R \omega \log\left[1+4\tau_R^2\omega^2\right]}{4 \tau_R^2 \omega^2}\right).
	\end{eqnarray}
The corresponding photo-emission rate is then given by 
	\begin{eqnarray}
	\frac{d\Gamma}{T^3dk}=-\frac{e^2\mathcal{A}_f|{\bf k|}}{e^{|k|/T}-1}\frac{\text{Im}\left[G^R_{1,1}(|{\bf k}|,|{\bf k}|)\right]}{\pi^2 T^3},
	\end{eqnarray}
where $\mathcal{A}_f$ denotes the charge and flavor degrees of freedom. For example, considering a 3-flavor system with u,d,s quarks and anti-quarks, we have $\mathcal{A}_f=4/3$. In the IR limit ($|{\bf k}|\rightarrow 0$), one finds
\begin{eqnarray}
\left(\frac{d\Gamma}{T^3dk}\right)_{|{\bf k}|/T\rightarrow 0}=\frac{ \mathcal{A}_f\chi e^2 \tau_R |{\bf k}|}{3 \pi ^2 T^2}+\mathcal{O}(|{\bf k}|^2/T^2)
\end{eqnarray}
and the corresponding electric conductivity reads
\begin{eqnarray}
\sigma=\frac{e^2\mathcal{A}_f\chi\tau_R}{3}.
\end{eqnarray}

We then make smooth interpolation between the emission rate from pQCD in the UV region and that from the RT approximation in the IR at $|{\bf k}|\approx 3T$ at $\mu=0$.
%by simply tuning $\tau_R$ at $\mu=0$ is unfortunately not possible due to the smaller rate from the RT approximation with $\mathcal{A}_f=4/3$.
For phenomenological purposes, we leave $\tau_R$ and $\mathcal{A}_f$ as  fitting parameters. By choosing $\tau_RT= 0.23$ and $\mathcal{A}_f=25.15$ for $\mu=0$ such that $\chi=T^2/12$, we are able to smoothly match the amplitude and slope of the rates from two approaches at $|{\bf k}|\approx 3T$. The interpolated emission rate is shown by the brown curve in Fig.\ref{interpolation_rate_2}. In the same figure, we also plot the perturbative rates intercepted by the linear extrapolations, 
\begin{eqnarray}
\left(\frac{d\Gamma}{T^3dk}\right)\approx \frac{\sigma|{\bf k}|}{\pi^2T^2},
\end{eqnarray}
at small $k/T$ given the electric conductivity from lattice results at different temperature. It turns out that the corresponding electric conductivity from the RT approximation yields $C_{em}^{-1}\sigma/T\approx 0.24$, which is close to lattice result around $T=1.4T_c$. Nonetheless, there exists a caveat that $\mathcal{A}_f=25.15$ ($\mathcal{A}_f=4/3$ in QCD) corresponds to unrealistically large flavor and charge degrees of freedom. On the other hand, as illustrated in Fig.\ref{RT_comparison}, it is enlightening to note that the emission rates fully obtained from RT approximations also give rise to the blue-shift at strong coupling.    

In Fig.~\ref{fig.koverT_stat}, we study the statistical distribution of $k/T$ at which the QGP rate is evaluated when we convolute the QGP rate with a hydrodynamic medium. Here we include the blue-shift effect to the photon energy $k = p \cdot u$ (not to be confused with the blue-shift of emission rates at strong coupling), where $p$ denotes the photon momentum in the lab frame. The QGP photons within transverse momentum range of 0.5 to 1 GeV are dominant by the values of QGP rate at $k/T \sim 4$, close to the IR modified $k/T$ region. For the photons with $p_T > 1$ GeV, their relevant production rates are at larger $k/T$ values. We expect that they are less affected by our IR treatment. 
In Fig.~\ref{fig.int_spectra} and Fig.\ref{fig.int_flow}, we compare the spectra and flow of QGP photons and direct photons in RHIC and LHC energies from pQCD rates with distinct interpolations in low energy. It turns out that the deviations in spectra are invisible. Although the interpolated rates yield mild modifications at low $p_T$ for QGP photons $v_2$, their impacts on the total direct photon anisotropic flow are negligible as well. 

%\bibliography{ref}
%\input{A_main_text.bbl}
\bibliography{A_main_text.bbl}

\end{document}